\documentclass[12pt]{article}
\usepackage{a4wide}
\usepackage{amstext,amsfonts,amsmath,theorem}
\usepackage{graphicx}

\usepackage{verbatim}
\makeatletter \@addtoreset{equation}{section}
\@addtoreset{table}{section}

\makeatother

        {\theorembodyfont{\rmfamily}
\newtheorem{prop}{Proposition}}

        {\theorembodyfont{\rmfamily}
\newtheorem{Th}{Theorem}}

{\end{list}}

\begin{document}
\newcounter{bean}

\thispagestyle{empty}


\begin{flushright}
FTUV--02/1231\quad IFIC--02/55
\\
31-Dec-02
\\[1cm]
\end{flushright}


\begin{center}

\begin{Large}
{\bf Generating Lie and gauge free differential (super)algebras by
expanding Maurer-Cartan forms and Chern-Simons supergravity}
\end{Large}
\vskip 1cm

\begin{large}
J.~A.~de~Azc\'arraga$^{a,1}$, J.~M. Izquierdo$^{b,1}$,
M.~Pic\'on$^{a,1}$ and O.~Varela$^{a,}$\footnote{
j.a.de.azcarraga@ific.uv.es, izquierd@fta.uva.es,
moises.picon@ific.uv.es, oscar.varela@ific.uv.es}
\end{large}
\vspace*{0.6cm}

\begin{it}
$a$ Departamento de F\'{\i}sica Te\'orica, Facultad de
F\'{\i}sica,
Universidad de Valencia\\
and IFIC, Centro Mixto Universidad de Valencia--CSIC,
\\
E--46100 Burjassot (Valencia), Spain
\\
$b$ Departamento de F\'{\i}sica Te\'orica, Universidad de
Valladolid
\\
E--47011 Valladolid, Spain
\\[0.4cm]
\end{it}
\end{center}

\begin{abstract}
We study how to generate new Lie algebras $\mathcal{G}(N_0,\ldots,
N_p,\ldots,N_n)$ from a given one $\mathcal{G}$. The (order by
order) method consists in expanding its Maurer-Cartan one-forms in
powers of a real parameter $\lambda$ which rescales the
coordinates of the Lie (super)group $G$, $g^{i_p} \rightarrow
\lambda^p g^{i_p}$, in a way subordinated to the  splitting of
$\mathcal{G}$ as a sum $V_0 \oplus \cdots \oplus V_p \oplus \cdots
\oplus V_n$ of vector subspaces. We also show that, under certain
conditions, one of the obtained algebras may correspond to a
generalized \.In\"on\"u-Wigner contraction in the sense of
Weimar-Woods, but not in general. The method is used to derive the
M-theory superalgebra, including its Lorentz part, from
$osp(1|32)$. It is also extended to include gauge free
differential (super)algebras and Chern-Simons theories, and then
applied to $D=3$ CS supergravity.
\end{abstract}

\section{Introduction and motivation: four methods to derive
new Lie algebras from given ones} \label{introd}

          The relation of given Lie algebras (and groups) among themselves,
and specially the derivation of new algebras from them, is a
problem of great interest in mathematics and physics, where it
goes back to the old problem of mixing of symmetries and to the
advent of supersymmetry itself, the only non-trivial way of
enlarging spacetime symmetries (see, respectively, \cite{Dyson}
and \cite{collsusy} and the papers reprinted therein). Setting
aside the trivial problem of finding whether a Lie algebra is a
subalgebra of another one there are, essentially, three different
ways of relating and/or obtaining new algebras from given ones.

          The {\it first} one is the {\it contraction} procedure
\cite{Seg51,IW53,Sal61}. In its \.In\"on\"u and Wigner (IW) simple
form \cite{IW53}, the contraction  $\mathcal{G}_c$ of a Lie
algebra $\mathcal G$ is performed with respect to a subalgebra
$\mathcal{L}_0$ by rescaling the basis generators of the coset
$\mathcal{G} / \mathcal{L}_0$ by means of a parameter, and then by
taking a singular limit for this parameter. The generators in
$\mathcal{G} / \mathcal{L}_0$ become abelian in the contracted
algebra $\mathcal{G}_c$, and the subalgebra $\mathcal{L}_0 \subset
\mathcal{G}_c$ acts on them. As a result, $\mathcal{G}_c$ has a
semidirect structure, and the abelian generators determine an
ideal of $\mathcal{G}_c$; obviously, $\mathcal{G}_c$ has the same
dimension as $\mathcal{G}$. The contraction process has well known
physical applications as {\it e.g.}, in understanding the
non-relativistic limit from a group theoretical point of view, or
to explain the appearance of dimensionful generators when the
original algebra $\mathcal{G}$ is semisimple (and hence with
dimensionless generators). This is achieved by using a
dimensionful contraction parameter, as in the derivation of the
Poincar\'e group from the de Sitter groups (there, the parameter
is the radius $R$ of the universe, and the limit is $R \rightarrow
\infty$). There have been many discussions and variations of the
IW contraction procedure (see \cite{AC79, CelTar, Lord85,MonPat91,
HMOS94, Wei:00} to name a few), but all of them have in common
that $\mathcal{G}$ and $\mathcal{G}_c$ have, necessarily, the same
dimension as vector spaces. The contraction process has also been
considered for `quantum' algebras (see {\it e.g.}, \cite{CGST92}).

           The {\it second} procedure is the {\it deformation} of
algebras, and Lie algebras in particular
\cite{Gerst64,NijRich66,NijRich67b,Rich67} (see also
\cite{Le67,Her70,Gil72}), which allows us to obtain algebras {\it
close}, but not isomorphic, to a given one.  This leads to the
important notion of rigidity \cite{Gerst64,NijRich66, Rich67} (or
physical stability): an algebra is called {\it rigid} when any
attempt to deform it leads to an equivalent (isomorphic) one. From
a physical point of view, the deformation process is essentially
the inverse to the contraction one (see \cite{Le67} and the second
ref. in \cite{Wei:00}), and the dimensions of the original and
deformed Lie algebras are again the same. For instance, the
Poincar\'e algebra is not rigid, but the de Sitter algebras, being
semisimple, have trivial second cohomology group by the Whitehead
lemma and, as a result, they are rigid. One may also consider the
Poincar\'e algebra as a deformation of the Galilei algebra, so
that this deformation may be read as a group theoretical
prediction of relativity. Thus, the mathematical deformation may
be physically considered as a tool for developing a physical
theory from another pre-existing one. Quantization itself may also
be looked at as a deformation (see \cite{Mo49,FLS76,Vey75}), the
classical limit being the contraction limit $\hbar \rightarrow 0$.

           A {\it third} procedure to obtain new Lie algebras is the
{\it extension} $\tilde{\mathcal{G}}$ of an algebra $\mathcal{G}$
by another one $\mathcal A$ (for details and references see {\it
e.g.} \cite{AI95}). The extended algebra $\tilde{\mathcal{G}}$
contains $\mathcal A$ as an ideal and $\tilde{\mathcal G}/\mathcal
A \approx \mathcal{G}$, but $\mathcal{G}$ is not necessarily a
subalgebra of $\tilde{\mathcal{G}}$. The data of the extension
problem is $\mathcal{G}$, $\mathcal A$ and an action of $\mathcal
G$ on $\mathcal A$. When $\mathcal A$ is abelian the problem
always has a solution, the semidirect sum $\tilde{\mathcal G}=
{\mathcal A} \supset \!\!\!\!\!\! \raisebox{1.5pt} {{\tiny +}}
\;\;\mathcal{G}$ (in which case $\mathcal{G}$ is a subalgebra of
$\tilde{\mathcal G}$), but in the general case there may be an
obstruction to the extension. Since, for an extension
$\tilde{\mathcal G}$, $\tilde{\mathcal G}/\mathcal A \approx
\mathcal G$ always, $\textrm{dim} \, \tilde{\mathcal G} =
\textrm{dim} \, \mathcal G + \textrm{dim} \, \mathcal A$. Thus,
once more, the dimension of the resulting algebra is equal to the
the total number of generators in the algebras involved in
obtaining the new one (here, the extension $\tilde{\mathcal G}$).

           The extension and deformation procedure are both
directly governed by various aspects of the cohomology of Lie
algebras. For instance, the existence of non-trivial {\it central}
extensions of $\mathcal G$ by an abelian algebra $\mathcal{A}$
depends solely on the non-triviality of the second cohomology
group $H_0^2(\mathcal{G},\mathcal{A})$. In this case there exist
non-trivial two-cocycles (the constant that accompanies them may
also play a dimensionally fundamental role). But cohomology also
plays a subtle role in the case of the contraction process since,
in general, the contraction generates cohomology. This explains
{\it e.g.}, how the 11-dimensional Galilei group (which is a
non-trivial central extension of the ordinary Galilei group by
$U(1)$) may be obtained from the {\it direct} product of the
Poincar\'e group by $U(1)$. This is possible because central
extension two-coboundaries, which correspond to trivial, direct
products, may become non-trivial two-cocycles in the contraction
limit \cite{Sal61,AA85}; other examples of this mechanism will be
mentioned in Sec.~\ref{sg}. In the case of deformations, a
sufficient condition for the rigidity of $\mathcal{G}$ is the
vanishing of the second cohomology group, $H^2(\mathcal G ,
\mathcal G)=0$ (hence, all semisimple algebras are rigid). The
elements of $H^2(\mathcal G , \mathcal G) \neq 0$ describe
infinitesimal (first order) deformations, which are integrable
into a one-parameter family of deformations if $H^3(\mathcal G ,
\mathcal G)=0$ \cite{Gerst64, NijRich67b}.

           These procedures can be extended to {\it super} or
$\mathbb{Z}_2$-graded Lie algebras, with even (bosonic) and odd
(fermionic) generators, by taking into account the specific
properties of the Grassmann variables. The cohomology of
superalgebras was briefly discussed first in \cite{Lei75} (see
also \cite{d'AFR80,Cd'AF91,SZ98}), and is especially important in the
context of supersymmetric theories (recent papers on superalgebra
extensions are \cite{SZ98,AMR,FL02}). Deformations of superalgebras
have also been considered (see {\it e.g.}, \cite{Bi86,SZ98}); for
instance, it may be seen  that $osp(1|4)$ is the only deformation of the
$D$=4, $N$=1 superPoincar\'e algebra \cite{Bi86}. One of the reasons
for the interest of superalgebra cohomology is its relevance
in the construction of actions for supersymmetric extended objects.
In particular, the generalization to superalgebras
of the Chevalley-Eilenberg approach \cite{CE48}
to Lie algebra cohomology is especially important in the
construction of the Wess-Zumino (WZ) terms that appear in the
superbrane actions \cite{AT89}. These terms may be shown to be
related to extensions of supersymmetry in various spacetime
dimensions. Indeed, it may well be that a better description of
supersymmetric extended objects requires that ordinary superspace
be enlarged with additional coordinates (beyond the standard
$(x,\theta)$ ones) following a fields/extended superspace
variables democracy principle (see \cite{CAIPB00} and references
therein). In fact, many of the spacetime supersymmetry algebras
(as the `M-algebra' \cite{To}\cite{AF82,vHvP82,Bars97,Se97}), and
their associated enlarged superspaces, may be considered as
algebra/group extensions (see \cite{CAIPB00} and references
therein), containing central (and non-central) generators. On the
deformation side, one may also apply the algebraic rigidity
criterion to superspace \cite{Chr01}, since it is given by a group
extension \cite{AA85b} and some extensions may also be viewed as
examples of deformations.

Nevertheless, we may ask ourselves whether the three above
procedures are all that may be useful in finding and discussing
the underlying symmetry structure of supersymmetric theories and
their interrelations, particularly when these include non-flat
geometries as $AdS$ ones. Motivated by these considerations, we
want to explore in this paper another way to obtain new algebras of 
increasingly higher dimensions from a given
one $\mathcal G$. The idea, originally considered by
Hatsuda and Sakaguchi in \cite{HS} in
a less general context, consists in looking at the algebra ${\cal
G}$ as described by the Maurer-Cartan (MC) forms on the manifold
of its associated group $G$ and, after rescaling some of the group
parameters by a factor $\lambda$, in expanding the MC forms as
series in $\lambda$. We shall study the method here in general,
and discuss how it can be applied when the rescaling is
subordinated to a splitting of ${\cal G}$ into a sum of vector
subspaces. The resulting {\it fourth} procedure, the {\it
expansion} method to be described below, is different from the
three above albeit, when the algebra dimension does not change in
the process, it may lead to a simple IW or IW-generalized
contraction ({\it i.e.}, one that rescales the algebra generators
using different powers of $\lambda$) in the sense of Weimar-Woods
\cite{Wei:00}, but not always. Furthermore, the algebras to which
it leads have in general higher dimension than the original one,
in which case they cannot be related to it by any contraction or
deformation process. We shall term them {\it expanded algebras}.

The use of the MC forms to discuss new algebras and superalgebras
is specially convenient. It allows us to treat Lie (super)algebras
as a particular case of free differential algebras
\cite{Su77,Ni83,Cd'AF91} and, from a physical point of view, to
have ready the invariant forms that are used to construct actions.
In particular, the MC forms on superspace, either enlarged or not,
are essential in the formulation of the actions for supersymmetric
objects as already mentioned. In fact, there have been indications
that the method below may be used \cite{HS,HS02} in the
construction, starting from the superalgebra of the $AdS$
superstring \cite{MT98}, of a Lie superalgebra that is realized in
the context of the IIB D-string. The discussion may also be
relevant when considering the structure of all possible enlarged
new superspaces, and  in particular to see whether the method of
\cite{CAIPB00}, which corresponds to the third procedure above
(extension of superalgebras) exhausts the number of possible,
physically relevant, superspaces. As one might expect, we shall
conclude that the extension method leads to a greater variety of
(super)algebras and that the expansion procedure may be useful to
find and relate existing theories.

In this paper we shall restrict ourselves to giving the general
structure of the expansion method (Secs.~\ref{dos}-\ref{susy})
plus some immediate applications concerning the M-theory
superalgebra (Sec.~\ref{seis}), gauge free differential
(super)algebras (Sec.~\ref{gfda}) and Chern-Simons supergravity in
$D=3$ (Sec.~\ref{sg}), leaving further developments for a
forthcoming paper.

\section{Rescaling of the group parameters and the
expansion method} \label{dos}

Let $G$ be a Lie group, of local coordinates $g^i$, $i=1, \ldots
,r=\textrm{dim}\,G$. Let $\mathcal{G}$ be its Lie algebra
\footnote{Calligraphic ${\mathcal G}$, ${\mathcal L}$, ${\mathcal
W}$ will denote both the Lie algebras and their underlying vector
spaces; $V$, $W$ etc. will be used for vector spaces that are not
necessarily Lie algebras.} of basis $\{X_i \}$, which may be
realized by left-invariant (LI) generators $X_i(g)$ on the group
manifold. Let  ${\mathcal G}^*$ be the coalgebra, and let  $\{
\omega ^i (g) \}$, $i=1,\dots,r= \mathrm{dim} \, G$ be the basis
determined by the (dual, LI) Maurer-Cartan (MC) one-forms on $G$.
Then, when $[X_i,X_j]=c_{ij}^k X_k$, the MC equations read
\begin{equation} \label{eq:mc}
d\omega^k(g)=-\frac{1}{2}c_{ij}^k \omega^i(g) \wedge \omega^j(g)
\; , \quad i,j,k=1,\ldots,r \quad .
\end{equation}
\noindent

We wish to show in this section how we may obtain new algebras by
means of a redefinition $g^{l}\rightarrow \lambda g^l$ of some of
the group parameters and by looking at the power series expansion
in $\lambda$ of the resulting one-forms $\omega^{i}(g,\lambda)$.
Let $\theta$ be the LI canonical form on $G$,
\begin{equation}
        \theta(g)=g^{-1}dg=
e^{-g^i X_i}\, d e^{g^i X_i} \equiv \omega^i X_i \; .
\end{equation}
\noindent Since {\setlength\arraycolsep{2pt}
\begin{eqnarray}
e^{-A}\, d e^{A} & = & dA + \frac12 [dA,A]+\frac{1}{3!}[[dA,A],A]+
\frac{1}{4!}[[[dA,A],A],A]+\ldots
\nonumber \\
& = & dA + \sum_{n=1}^{\infty}
\frac{1}{(n+1)!}[\stackrel{n}{\ldots} [dA,A],\ldots,A], A] \; ,
\end{eqnarray}}
\noindent one obtains, for $A \equiv g^k X_k$, $dA =(dg^j)X_j$,
the expansion of $\theta(g)$ and of the MC forms $\omega^i(g)$ as
polynomials in the group coordinates $g^i$ :
\begin{equation}
\theta(g) =  \left[ \delta_j^i +\frac{1}{2!} c_{j k}^i g^k +
\frac{1}{3!} c_{j k_1}^{h_1}c_{h_1 k_2}^i g^{k_1} g^{k_2} +
        \frac{1}{4!} c_{j k_1}^{h_1}c_{h_1 k_2}^{h_2}c_{h_2 k_3}^{i}
g^{k_1} g^{k_2} g^{k_3}+ \ldots \right] dg^j X_i \quad ,
\label{eq:theta}
\end{equation}
\begin{equation} \label{eq:serie2}
\omega^i(g) =  \left[ \delta_j^i +\frac{1}{2!} c_{j k}^i g^k+
    \sum_{n=2}^{\infty}
\frac{1}{(n+1)!} c_{j k_1}^{h_1}c_{h_1 k_2}^{h_2} \ldots
        c_{h_{n-1} k_{n-1}}^{h_{n-1}}c_{h_{n-1} k_n}^{i}
g^{k_1} g^{k_2} \ldots g^{k_{n-1}} g^{k_n} \right] dg^j \quad .
\end{equation}

\noindent Looking at (\ref{eq:serie2}), it is evident that the
redefinition
\begin{equation}
g^l \rightarrow \lambda g^l
\end{equation}
of {\it some} coordinates $g^l$ will produce an expansion of the
MC one-forms $\omega^i(g, \lambda)$ as a sum of one-forms
$\omega^{i,\alpha}(g)$ on $G$ multiplied by the corresponding
powers $\lambda^\alpha$ of $\lambda$.

\subsection{The Lie algebras $\mathcal{G}(N)$ generated from
$\mathcal{G}=V_0\oplus V_1$}

Consider, as a first example, the splitting of ${\mathcal G}^*$
into the sum of two (arbitrary) vector subspaces,
\begin{equation} \label{split}
{\mathcal G}^*  = V_0^* \oplus V_1^* \; ,
\end{equation}
$V^*_0$, $V^*_1$ being generated by the MC forms
$\omega^{i_0}(g)$, $\omega^{i_1}(g)$ of ${\mathcal G}^*$ with
indexes corresponding, respectively, to the unmodified and
modified parameters,
\begin{equation} \label{eq:redefinition}
g^{i_{0}} \rightarrow  g^{i_{0}} \; , \; g^{i_{1}} \rightarrow
\lambda g^{i_{1}}  \quad,\quad i_0 \; (i_1) = 1, \ldots,
\textrm{dim} \, V_0 \; (\textrm{dim} \, V_1) \,.
\end{equation}
In general, the series of $\omega^{i_0}(g,\lambda) \in V_0^*$,
$\omega^{i_1}(g,\lambda) \in V_1^*$, will involve all powers of
$\lambda$,
\begin{equation} \label{eq:series}
\omega^{i_{ p}}(g,\lambda)  =  \sum_{\alpha=0}^{\infty}
\lambda^\alpha \omega^{i_{p},\alpha}(g) = \omega^{i_{ p},0}(g) +
\lambda \omega^{i_{ p},1}(g) + \lambda^2 \omega^{i_{ p},2}(g) +
\ldots \; ; \quad  p=0,1 \; ,
\end{equation}
$\omega^{i_p}(g,1)=\omega^{i_p}(g)$. We will see in the following
sections what restrictions on $\mathcal G$ make zero certain
coefficient one-forms $\omega^{i_{p},\alpha}$.

With the above notation, the MC equations (\ref{eq:mc}) for
$\mathcal G$
       can be rewritten as
\begin{equation} \label{eq:MC}
d\omega^{k_{ s}}= -\frac{1}{2} c_{i_{ p}j_{ q}}^{k_{ s}} \,
\omega^{i_{ p}} \wedge \omega^{j_{ q}} \quad (p,q,s=0,1)
\end{equation}
or, explicitly {\setlength\arraycolsep{2pt}
\begin{eqnarray}
\label{eq:MCzero} d\omega^{k_{ 0}} & = & -\frac{1}{2} c_{i_{ 0}j_{
0}}^{k_{ 0}} \omega^{i_{ 0}} \wedge \omega^{j_{ 0}} - c_{i_{ 0}j_{
1}}^{k_{ 0}} \omega^{i_{ 0}} \wedge \omega^{j_{ 1}} -\frac{1}{2}
c_{i_{ 1}j_{ 1}}^{k_{ 0}}
\omega^{i_{ 1}} \wedge \omega^{j_{ 1}}  \\
\label{eq:MCone} d\omega^{k_{ 1}} & = & -\frac{1}{2} c_{i_{ 0}j_{
0}}^{k_{ 1}} \omega^{i_{ 0}} \wedge \omega^{j_{ 0}} - c_{i_{ 0}j_{
1}}^{k_{ 1}} \omega^{i_{ 0}} \wedge \omega^{j_{ 1}} -\frac{1}{2}
c_{i_{ 1}j_{ 1}}^{k_{ 1}} \omega^{i_{ 1}} \wedge \omega^{j_{ 1}}
\; .
\end{eqnarray}}

\noindent Inserting now the expansions (\ref{eq:series}) into the
MC equations (\ref{eq:MC}) and using (\ref{eq:sumatorio}) in the
Appendix, the MC equations are expanded in powers of $\lambda$:
\begin{equation} \label{eq:MCexpanded}
\sum_{\alpha=0}^{\infty} \lambda^\alpha d\omega^{k_{ s}, \alpha}=
\sum_{\alpha=0}^{\infty} \lambda^\alpha \left[ -\frac{1}{2} c_{i_{
p}j_{ q}}^{k_{ s}} \sum_{\beta=0}^{\alpha} \omega^{i_{ p}, \beta}
\wedge \omega^{j_{ q}, \alpha-\beta} \right] \; .
\end{equation}
The equality of the two $\lambda$-polynomials in
(\ref{eq:MCexpanded}) requires the equality of the coefficients of
equal power $\lambda^{\alpha}$. This implies that the coefficient
one-forms $\omega^{i_{ p}, \alpha}$ in the expansions
(\ref{eq:series}) satisfy the identities:
\begin{equation} \label{eq:MCG}
d\omega^{k_{ s}, \alpha}= -\frac{1}{2} c_{i_{ p}j_{ q}}^{k_{ s}}
\sum_{\beta=0}^{\alpha} \omega^{i_{ p}, \beta} \wedge \omega^{j_{
q}, \alpha-\beta} \quad (p,q,s=0,1) \quad .
\end{equation}
We can rewrite (\ref{eq:MCG}) in the form
\begin{equation} \label{eq:cnts}
d\omega^{k_{ s}, \alpha}= -\frac{1}{2} C_{i_{ p},\beta \; j_{
q},\gamma}^{k_{ s},\alpha}\;
       \omega^{i_{ p}, \beta} \wedge
\omega^{j_{ q}, \gamma} \;, \;\;\; C_{i_{ p},\beta \; j_{
q},\gamma}^{k_{ s},\alpha}= \left\{
\begin{array}{lll} 0\,, &
\mathrm{if} \ \beta + \gamma \neq \alpha  \\
c_{i_{ p}j_{ q}}^{k_{s}}\,, & \mathrm{if} \ \beta + \gamma =
\alpha
\end{array} \right.\quad .
\end{equation}

We now ask ourselves whether we can use the expansion coefficients
$\omega^{k_{ 0}, \alpha}$, $\omega^{k_{ 1}, \beta}$ up to given
orders $N_{ 0} \geq 0$, $N_{1} \geq 0$, $\alpha=0,1,\ldots, N_0$,
$\beta=0,1,\ldots, N_1$, so that eq.~(\ref{eq:cnts}) (or (\ref{eq:MCG}))
determines the MC equations of a new Lie algebra.
The answer is given by the following

\begin{Th}
Let $\mathcal G$ be a Lie algebra, and ${\mathcal G} = V_0 \oplus
V_1$ (no subalgebra condition is assumed neither for $V_0$ or
$V_1$). Let $\{\omega^i \}$, $\{ \omega^{i_0} \}$, $\{
\omega^{i_1} \}$ ($i=1,\ldots, \textrm{dim} \,{\mathcal G}$,
$i_0=1,\ldots,\textrm{dim}\,V_0$, $i_1=1,\ldots, \textrm{dim}
\,V_1$) be, respectively, the bases of the ${\mathcal G}^*$,
$V_0^*$ and $V_1^*$ dual vector spaces. Then, the vector space
generated by
\begin{equation} \label{eq:largerbasis1}
\{ \omega^{i_{ 0}, 0}, \omega^{i_{ 0}, 1}, \ldots, \omega^{i_{ 0},
N}; \omega^{i_{ 1}, 0}, \omega^{i_{ 1}, 1}, \ldots, \omega^{i_{
1}, N} \} \; ,
\end{equation}
together with the MC eqs.~(\ref{eq:cnts}) for the structure
constants
\begin{equation} \label{eq:cnt2}
C_{i_{ p},\beta \; j_{ q},\gamma}^{k_{ s},\alpha}= \left\{
\begin{array}{lll}
0, & \mathrm{if} \ \beta + \gamma \neq \alpha  \\
c_{i_{ p}j_{ q}}^{k_{s}}, & \mathrm{if} \ \beta + \gamma = \alpha
\end{array} \right.
       \quad ( \alpha, \beta, \gamma = 1, \ldots ,N\;;\;p,q,s=0,1) \; ,
\end{equation}
determines a Lie algebra $\mathcal{G}(N)$ for each expansion order
$N \geq 0$  of dimension $\textrm{dim} \, \mathcal{G}(N) = (N+1)
\,  \textrm{dim} \, \mathcal{G}$.
\end{Th}

\noindent {\it Proof}.

\noindent Consider the one-forms
\begin{equation} \label{eq:lb1}
\{\omega^{i_0,\alpha_0}\,;\,\omega^{i_1,\alpha_1}\}= \{
\omega^{i_{ 0}, 0}, \omega^{i_{ 0}, 1}, \ldots, \omega^{i_{ 0},
N_{ 0}}; \omega^{i_{ 1}, 0}, \omega^{i_{ 1}, 1}, \ldots,
\omega^{i_{ 1}, N_{ 1}} \}
\end{equation}
where we have not assumed {\it a priori} the same range for the
expansions of the one-forms of $V_0^*$ and $V_1^*$. To see whether
the vector space  $V^*(N_{ 0}, N_{ 1})$ of basis (\ref{eq:lb1})
determines a Lie algebra $\mathcal{G}(N_0,N_1)$, it is sufficient
to check that a) that the exterior algebra generated by
(\ref{eq:lb1}) is closed\footnote{An algebra of forms closed under
$d$ defines in general a free differential algebra (FDA)
\cite{Su77,AF82,Ni83,Cd'AF91} (FDA's were called Cartan integrable
systems in \cite{AF82}). When the generating forms are one-forms,
the FDA corresponds to a Lie algebra.} under the exterior
derivative $d$ and that a) the Jacobi identities (JI) for
${\mathcal G}$ are satisfied.

To have closure under $d$ we need that the {\it r.h.s.} of eqs.
(\ref{eq:cnts}) does not contain one-forms that are not already
present in (\ref{eq:lb1}). Consider the forms
$\omega^{i_s,\beta_s}$, $s=0,1$ that contribute to $d \omega^{k_{
s},\alpha_s}$ up to order $\alpha=N_s$. Looking at
eqs.~(\ref{eq:MCG}) it follows trivially that
\begin{equation} \label{eq:N}
N_0=N_1 \qquad (= N) \; .
\end{equation}
\noindent To check the JI for $\mathcal{G}(N)$, it is sufficient
to see that $dd\omega^{k_s,\alpha}\equiv 0$ in (\ref{eq:cnts}) is
consistent with the definition of $C_{i_p,\beta \;
j_q,\gamma}^{k_s,\alpha}\,$. Eq.~(\ref{eq:cnts}) gives
\begin{equation} \label{eq:jacobi}
0 = C_{i_p,\beta\;j_q,\gamma}^{k_s,\alpha} C_{l_t,\rho\;
m_u,\sigma}^{i_p,\beta}\omega^{j_q,\gamma}\wedge
\omega^{l_t,\rho}\wedge\omega^{m_u,\sigma} \quad
(\alpha,\beta,\gamma,\rho,\sigma=1,\ldots,N)\;,
\end{equation}
\noindent which implies
\begin{equation}\label{eq:bjacobi}
C_{i_p,\beta\;[j_q,\gamma}^{k_s,\alpha} C_{l_t,\rho\;
m_u,\sigma]}^{i_p,\beta}=0\; .
\end{equation}
\noindent Now, on account of definition (\ref{eq:cnt2}), the terms
in the {\it l.h.s.} above are either zero (when
$\alpha\not=\gamma+\rho+\sigma$) or give zero due to the JI for
${\mathcal G}$, $c_{i_{ p} [ j_{ q}}^{k_s} c_{l_t m_u ]}^{i_p} =0
$. Thus, the $C_{i_p,\beta \;j_q,\gamma}^{k_s,\alpha}\,$ satisfy
the JI (\ref{eq:bjacobi}) and define the Lie algebra
$\mathcal{G}(N,N) \equiv \mathcal{G}(N)$, {\it q.e.d.}\\

Explicitly, the resulting algebras for the first orders are:\\[6pt]
\noindent $N=0\;,\;{\cal G}$(0):
\begin{equation} \label{eq:Nzero}
d\omega^{k_{ s}, 0}= -\frac{1}{2} c_{i_{ p}j_{ q}}^{k_{ s}}
\omega^{i_{ p}, 0} \wedge \omega^{j_{ q}, 0} \quad (p,q,s=0,1) \;
,
\end{equation}
\noindent{\it i.e.}, $\mathcal{G}(0)$ reproduces the original
algebra $\mathcal{G}$.
\\[6pt]
\noindent $N=1\;,\;{\cal G}$(1): {\setlength\arraycolsep{2pt}
\begin{eqnarray} \label{eq:N10}
d\omega^{k_{ s}, 0}&=& -\frac{1}{2}  c_{i_{ p}j_{ q}}^{k_{
s}} \omega^{i_{ p}, 0} \wedge \omega^{j_{ q}, 0}  \; ,  \\
\label{eq:N11} d\omega^{k_{ s}, 1}&=& -c_{i_{ p}j_{ q}}^{k_{ s}}
\omega^{i_{ p}, 0} \wedge \omega^{j_{ q}, 1} \quad (p,q,s=0,1) \;
\end{eqnarray}} \\[-10pt]
\noindent $N=2\;,\;{\cal G}$(2): {\setlength\arraycolsep{0.5pt}
\begin{eqnarray} \label{eq:N20}
\!\!\!\!\!\!\!\!\!\!\!\!\! d\omega^{k_{ s}, 0}&=& -\frac{1}{2}
c_{i_{p}j_{ q}}^{k_{
s}} \omega^{i_{ p}, 0} \wedge \omega^{j_{ q}, 0} \; , \\
\label{eq:N21} \!\!\!\!\!\!\!\!\!\!\!\!\! d\omega^{k_{ s}, 1}&=& -
c_{i_{ p}j_{ q}}^{k_{s}} \omega^{i_{ p}, 0} \wedge \omega^{j_{ q},
1}
\; ,\\
\label{eq:N22} \!\!\!\!\!\!\!\!\!\!\!\!\! d\omega^{k_{ s}, 2}&=&
-c_{i_{ p}j_{ q}}^{k_{ s}} \omega^{i_{ p}, 0} \wedge \omega^{j_{
q}, 2} -\frac{1}{2} c_{i_{ p}j_{q}}^{k_{ s}} \omega^{i_{ p}, 1}
\wedge \omega^{j_{ q}, 1} \;\; (p,q,s=0,1) \quad .
\end{eqnarray}}

\noindent {\it Remark}. Since $\omega^{i_{ p}, 0}(g) \neq
\omega^{i_{ p}}(g)$, one might wonder {\it e.g.} how the MC eqs.
for $\mathcal{G}(0)=\mathcal{G}$ can be satisfied by $\omega^{i_{
p},0}(g)$. The $\mathrm{dim}\,\mathcal{G}$ MC forms
$\omega^{i_{p}}(g)$ are LI forms on the group manifold $G$ of
$\mathcal{G}$. The $(N+1)\mathrm{dim}\,\mathcal{G}$
$\omega^{i_{p}, \alpha}(g)$ ($\alpha=0,1,\ldots,N$) determined by
the expansions (\ref{eq:series})  are also one-forms on $G$, but
they are no longer LI under $G$-translations. They cannot be,
since there are only $\mathrm{dim} \, G=r$ linearly independent MC
forms on $G$. Nevertheless, eqs. (\ref{eq:cnts}) determine the MC
relations that will be satisfied by the MC forms on the manifold
of the {\it higher dimensional} group $G(N)$ associated with
$\mathcal{G}(N)$. These MC forms on $G(N)$ will depend on the
$(N+1)\mathrm{dim} \, \mathcal{G}(N)$ coordinates of $G(N)$
associated with the generators (forms) $X_{i_p, \alpha}$
($\omega^{i_p,\alpha}$) that determine $\mathcal{G}(N)$
(${\mathcal G}^*(N)$).

\subsection{Structure of the Lie algebras $\mathcal{G}(N)$}
\label{s:structure}

Let $V_{p,\alpha}$ be, at each order $\alpha=0,1,\ldots, N$, the
vector space spanned by the generators $X_{{i_p},\alpha}$,
$p=0,1$; clearly, $V_{p,\alpha} \approx V_{p}$. Let
\begin{equation} \label{def:W}
W_{\alpha}=V_{0,\alpha} \oplus V_{1,\alpha} \quad , \quad
{\mathcal G}(N) = \bigoplus_{\alpha=0}^N W_{\alpha} \quad .
\end {equation}
We first notice that ${\mathcal G}(N-1)$ is a vector subspace of
${\mathcal G}(N)$, but not a subalgebra for $N \geq 2$. Indeed,
for $N \geq 2$ there always exist $\alpha,\beta \leq N-1$ such
that $\alpha+\beta=N$. Denoting by $C_{i_p,\alpha
\;j_q,\beta}^{(N)\,k_s,\gamma}$ and $C_{i_p,\alpha
\;j_q,\beta}^{(N-1)\,k_s,\gamma}$ the structure constants of
$\mathcal{G}(N)$ and $\mathcal{G}(N-1)$ respectively, one sees
that, for $\alpha+\beta=N$, $C_{i_p,\alpha \;
j_q,\beta}^{(N-1)\,k_s,\alpha+\beta}=0$ in $\mathcal{G}(N-1)$
(since $\alpha+\beta > N-1$) while, in general, $C_{i_p,\alpha \;
j_q,\beta}^{(N)\,k_s,\alpha+\beta} \neq 0$ in $\mathcal{G}(N)$. In
other words, $\mathcal{G}(N-1)$ is not a subalgebra of
$\mathcal{G}(N)$ because the structure constants for the elements
of the various subspaces $V_{p,\alpha}$ depend on $N$ and they are
different, in general, for $\mathcal{G}(N-1)$ and
$\mathcal{G}(N)$. Likewise, $\mathcal{G}(M)$ for $1 \leq M < N$ is
not a subalgebra of $\mathcal{G}(N)$.

We show in this section that the Lie algebras $\mathcal{G}(N)$
have a Lie algebra extension structure for $N \geq 1$.

\begin{prop} \label{prop:ext}
The Lie algebra $\mathcal{G}(0)$ is a subalgebra of
$\mathcal{G}(N)$, for all $N \geq 0$. For $N \geq 1$, $W_N$ is an
abelian ideal $\mathcal{W}_N \subset \mathcal{G}(N)$ and
$\mathcal{G}(N) / \mathcal{W}_N = \mathcal{G}(N-1)$ {\it i.e.},
$\mathcal{G}(N)$ is an extension of $\mathcal{G}(N-1)$ by
$\mathcal{W}_N$ which is not semidirect for $N \ge 2$.
\end{prop}

\noindent {\it Proof}.

\noindent $\mathcal{G}(0) \subset \mathcal{G}(N)$ is a subalgebra
by construction, since $C_{i_p,0 \; j_q,0}^{(N)\,k_s,\alpha}=0$,
$\alpha=1,\ldots,N$, by eq.~(\ref{eq:cnts}).

       For the second part, notice that, since $\alpha+N > N$ for $\alpha
\neq 0$, $[W_{\alpha}, \, \mathcal{W}_N] =0 $; in particular,
$\mathcal{W}_N$ is an abelian subalgebra. Furthermore
$[\mathcal{W}_0, \, \mathcal{W}_N] \subset \mathcal{W}_N$, so that
$\mathcal{W}_N$ is an ideal of $\mathcal{G}(N)$. Now, the vector
space $\mathcal{G}(N) / W_N$ is isomorphic to $\mathcal{G}(N-1)$.
$\mathcal{G}(N-1)$ is a Lie algebra the MC equations of which are
(\ref{eq:cnts}), and $\mathcal{G}(N) / \mathcal{W}_N \approx
\mathcal{G}(N-1)$. Since $\mathcal{G}(N-1)$ is not a subalgebra of
$\mathcal{G}(N)$ for $N\geq 2$, the extension is not semidirect,
{\it q.e.d.}

\subsection {The limiting cases $V_0=0,V_1=V$ and $V_0=V,V_1=0$}
When $V_1=V$, all the group parameters are modified by
(\ref{eq:redefinition}). In this case $\mathcal{G}(0)$ is the
trivial $\mathcal{G}(0)=0$ subalgebra of $\mathcal{G}(N)$. The
first order $N=1$, $\omega^{i_1,1}=dg^{i_1}$, corresponds to an
abelian algebra with the same dimension as $\mathcal{G}$ (in fact,
$\mathcal{G}(1)$ is the IW contraction of $\mathcal{G}$ with
respect to the trivial $V_0=0$ subalgebra). For $N \ge 2 $ we will
have extensions with the structure in Prop. \ref{prop:ext}.

For the other limiting case, $V_1=0$, there is obviously no
expansion and we have $\mathcal{G}(0)=\mathcal{G}$.

\section{The case in which $V_0$ is a subalgebra $\mathcal{L}_0\subset
\mathcal{G}$} \label{tres}

Let ${\mathcal G} = V_{ 0} \oplus V_{ 1}$ as before, where now
$V_0$ is a subalgebra $\mathcal{L}_0$ of $\mathcal{G}$. Then,
\begin{equation} \label{eq:subalgebra}
c_{i_0 j_0}^{k_1} = 0 \quad \quad (i_p = 1, \ldots, \textrm{dim}
\, V_p \, , \; p=0,1) \quad ,
\end{equation}
and the  basis one-forms $\omega^{i_0}$ are associated with the
(sub)group parameters $g^{i_0}$ unmodified under the rescaling
(\ref{eq:redefinition}). The MC equations for $\mathcal{G}$ become
{\setlength\arraycolsep{2pt}
\begin{eqnarray}
\label{eq:MCzero1} d\omega^{k_{0}} & = & -\frac{1}{2}
c_{i_{0}j_{0}}^{k_{0}} \omega^{i_{0}} \wedge \omega^{j_{0}} -
c_{i_{0}j_{1}}^{k_{0}} \omega^{i_{0}} \wedge \omega^{j_{1}}
-\frac{1}{2} c_{i_{1}j_{1}}^{k_{0}}
\omega^{i_{1}} \wedge \omega^{j_{1}} \quad , \\
\label{eq:MCone1} d\omega^{k_{1}} & = & - c_{i_{0}j_{1}}^{k_{1}}
\omega^{i_{0}} \wedge \omega^{j_{1}} -\frac{1}{2}
c_{i_{1}j_{1}}^{k_{1}} \omega^{i_{1}} \wedge \omega^{j_{1}} \; .
\end{eqnarray}}

Using (\ref{eq:subalgebra}) in eq. (\ref{eq:serie2}), one finds
that the expansions of $\omega^{i_0}(g,\lambda)$
($\omega^{i_1}(g,\lambda)$)  start with the power  $\lambda^0$
($\lambda^1$): {\setlength\arraycolsep{2pt}
\begin{eqnarray} \label{eq:serieszero}
\omega^{i_{0}}(g,\lambda) & = & \sum_{\alpha=0}^{\infty}
\lambda^\alpha \omega^{i_{0},\alpha}(g) = \omega^{i_{0},0}(g) +
\lambda \omega^{i_{0},1}(g) + \lambda^2 \omega^{i_{0},2}(g) +
\ldots  \\ \label{eq:seriesone} \omega^{i_{1}}(g,\lambda) & = &
\sum_{\alpha=1}^{\infty} \lambda^\alpha \omega^{i_{1},\alpha}(g) =
\lambda \omega^{i_{1},1}(g) + \lambda^2 \omega^{i_{1},2}(g) +
\lambda^3 \omega^{i_{1},3}(g) + \ldots  \; .
\end{eqnarray}}

\noindent Inserting them into the MC equations (\ref{eq:MCzero1})
and (\ref{eq:MCone1})
    and using eq.~(\ref{eq:sumatorio}) when the double sums begin with
$(0,0)$, $(0,1)$ and $(1,1)$, we get {\setlength\arraycolsep{2pt}
\begin{eqnarray}
\label{eq:MCins0} \sum_{\alpha=0}^{\infty} \lambda^\alpha
d\omega^{k_{0}, \alpha} & = & -\frac{1}{2} c_{i_{0}j_{0}}^{k_{0}}
\omega^{i_{0}, 0} \wedge \omega^{j_{0}, 0} + \lambda \left[
-c_{i_{0}j_{0}}^{k_{0}}
       \omega^{i_{0}, 0} \wedge \omega^{j_{0}, 1}
-c_{i_{0}j_{1}}^{k_{0}}
       \omega^{i_{0}, 0} \wedge \omega^{j_{1}, 1} \right]+
\nonumber \\
& + &  \sum_{\alpha=2}^{\infty} \lambda^\alpha \left[ -\frac{1}{2}
c_{i_{0}j_{0}}^{k_{0}} \sum_{\beta=0}^{\alpha} \omega^{i_{0},
\beta} \wedge \omega^{j_{0}, \alpha-\beta} -c_{i_{0}j_{1}}^{k_{0}}
\sum_{\beta=0}^{\alpha-1} \omega^{i_{0}, \beta} \wedge
\omega^{j_{1}, \alpha-\beta} - \right.
\nonumber  \\
&-& \left. \frac{1}{2} c_{i_{1}j_{1}}^{k_{0}}
       \sum_{\beta=1}^{\alpha-1} \omega^{i_{1}, \beta} \wedge
\omega^{j_{1}, \alpha-\beta} \right] \quad , \\
\label{eq:MCins1} \sum_{\alpha=1}^{\infty} \lambda^\alpha
d\omega^{k_{1}, \alpha} & = & -\lambda c_{i_{0}j_{1}}^{k_{1}}
       \omega^{i_{0}, 0} \wedge \omega^{j_{1}, 1}+
\nonumber  \\
& + & \sum_{\alpha=2}^{\infty} \lambda^\alpha \left[
-c_{i_{0}j_{1}}^{k_{1}} \sum_{\beta=0}^{\alpha-1} \omega^{i_{0},
\beta} \wedge \omega^{j_{1}, \alpha-\beta} -\frac{1}{2}
c_{i_{1}j_{1}}^{k_{1}}
       \sum_{\beta=1}^{\alpha-1} \omega^{i_{1}, \beta} \wedge
\omega^{j_{1}, \alpha-\beta} \right] \, .
\end{eqnarray}}

\noindent Again, the equality of the coefficients of equal power
$\lambda^\alpha$ in
(\ref{eq:MCins0}), (\ref{eq:MCins1}) leads to the equalities: \\[5pt]
$\alpha=0$:
\begin{equation} \label{eq:00}
d\omega^{k_{0}, 0} = -\frac{1}{2} c_{i_{0}j_{0}}^{k_{0}}
\omega^{i_{0}, 0} \wedge \omega^{j_{0}, 0} \quad ;
\end{equation}
$\alpha=1$: {\setlength\arraycolsep{2pt}
\begin{eqnarray}
\label{eq:01} d\omega^{k_{0}, 1} & = & -c_{i_{0}j_{0}}^{k_{0}}
       \omega^{i_{0}, 0} \wedge \omega^{j_{0}, 1}
-c_{i_{0}j_{1}}^{k_{0}}
       \omega^{i_{0}, 0} \wedge \omega^{j_{1}, 1} \quad , \\
\label{eq:11} d\omega^{k_{1}, 1} & = & -c_{i_{0}j_{1}}^{k_{1}}
       \omega^{i_{0}, 0} \wedge \omega^{j_{1}, 1} \quad ;
\end{eqnarray}}

\noindent $\alpha \geq 2$: {\setlength\arraycolsep{0pt}
\begin{eqnarray}
\label{eq:0r} \!\!d\omega^{k_{0}, \alpha} & = & -\frac{1}{2}
c_{i_{0}j_{0}}^{k_{0}} \sum_{\beta=0}^{\alpha} \omega^{i_{0},
\beta} \wedge \omega^{j_{0}, \alpha-\beta} -c_{i_{0}j_{1}}^{k_{0}}
\sum_{\beta=0}^{\alpha-1} \omega^{i_{0}, \beta} \wedge
\omega^{j_{1}, \alpha-\beta} -\frac{1}{2} c_{i_{1}j_{1}}^{k_{0}}
       \sum_{\beta=1}^{\alpha-1} \omega^{i_{1}, \beta} \wedge
\omega^{j_{1}, \alpha-\beta} \, , \quad \quad \;\;\; \\
\label{eq:1r} d\omega^{k_{1}, \alpha} & = &
-c_{i_{0}j_{1}}^{k_{1}} \sum_{\beta=0}^{\alpha-1} \omega^{i_{0},
\beta} \wedge \omega^{j_{1}, \alpha-\beta} -\frac{1}{2}
c_{i_{1}j_{1}}^{k_{1}}
       \sum_{\beta=1}^{\alpha-1} \omega^{i_{1}, \beta} \wedge
\omega^{j_{1}, \alpha-\beta} \, . \quad\quad \;
\end{eqnarray}}

To allow for a different range in the orders $\alpha$ of each
$\omega^{i_p, \alpha}$, we now denote the coefficient one-forms in
(\ref{eq:serieszero}) ((\ref{eq:seriesone})) $\omega^{i_0 ,
\alpha_0}$ ($\omega^{i_1 , \alpha_1}$), $\alpha_0=0,1,\ldots,N_0$
($\alpha_1=1,2,\ldots,N_1$). With this notation, the above
relations take the generic form
\begin{equation} \label{eq:MCsub}
d\omega^{k_{ s}, \alpha_s}= -\frac{1}{2} C_{i_{ p},\beta_p \;
j_{q},\gamma_q}^{k_s,\alpha_s}\; \omega^{i_{ p}, \beta_p} \wedge
\omega^{j_{ q}, \gamma_q} \quad ,
\end{equation}
where
\begin{equation} \label{eq:Csub}
C_{i_{ p},\alpha_p \; j_{ q},\alpha_q}^{k_{ s},\alpha_s}= \left\{
\begin{array}{lll} 0, &
\mathrm{if} \ \beta_p + \gamma_q \neq \alpha_s  \\
c_{i_{ p}j_{ q}}^{k_{s}}, & \mathrm{if} \ \beta_p + \gamma_q =
\alpha_s  \end{array} \right. \quad \begin{array}{l}
p,q,s=0,1 \\
i_{p,q,s}=1,2,\ldots, \textrm{dim} \, V_{p,q,s} \\
\alpha_0,\beta_0,\gamma_0=0,1, \ldots, N_0 \\
\alpha_1,\beta_1,\gamma_1=1,2, \ldots, N_1 \;  .
\end{array}
\end{equation}

As in the preceding case, we now ask ourselves whether the
expansion coefficients $\omega^{k_{0}, \alpha_0}$, $\omega^{k_{1},
\alpha_1}$ up to a given order $N_0,N_1$ determine the MC
equations (\ref{eq:MCsub}) of a new Lie algebra
$\mathcal{G}(N_0,N_1)$. It is obvious from (\ref{eq:00}) that the
zero order of the expansion in $\lambda$ corresponds to
$N_0=0=N_1$ (omitting all $\omega^{i_1,\alpha_1}$ and thus allowing
$N_1$ to be zero), and that ${\mathcal G}(0,0)={\mathcal L}_0$. It is
seen directly that the terms up to first order give two
possibilities: ${\mathcal G}(0,1)$ (eqs. (\ref{eq:00}),
(\ref{eq:11}) for $\omega^{k_0,0}\,, \omega^{k_1,1}$) and
${\mathcal G}(1,1)$ (eqs. (\ref{eq:00}), (\ref{eq:01}),
(\ref{eq:11}) for
$\omega^{k_0,0}\,,\omega^{k_1,1}\,,\omega^{k_0,1}$). Thus, we see
that now (and due to (\ref{eq:subalgebra})) one does not need to
retain {\it all} $\omega^{i_p,\alpha_p}$ up to a given order to
obtain a Lie algebra. To look at the general $N_0\geq 0, N_1\geq
1$ case, consider the vector space $V^*(N_{0}, N_{1})$, generated
by
\begin{equation} \label{eq:largerbasis}
\{ \omega^{i_0, \alpha_0}\,;\, \omega^{i_1, \alpha_1} \} = \{
\omega^{i_{0}, 0}, \omega^{i_{0}, 1}, \omega^{i_{0}, 2}, \ldots,
\omega^{i_{0}, N_{0}}; \omega^{i_{1}, 1}, \omega^{i_{1}, 2},
\ldots, \omega^{i_{1}, N_{1}} \} \; .
\end{equation}
To see that it determines a Lie algebra ${\mathcal G}(N_0,N_1)$ of
dimension
\begin{equation} \label{eq:dimsub}
\textrm{dim} \, \mathcal{G}(N_0,N_1)= (N_0+1) \, \textrm{dim} \,
V_0 + N_1 \textrm{dim} \, V_1 \; ,
\end{equation}
we first notice that the JI in $\mathcal{G}(N_{0}, N_{1})$ will
follow from the JI in $\mathcal{G}$. To find the conditions that
$N_{0}$ and $N_{1}$ must satisfy to have closure under $d$, we
look at the orders $\beta_p$ of the forms $\omega^{i_{p},\beta_p}$
that appear in the expression (\ref{eq:MCsub}) of $d
\omega^{k_{s},\alpha_s}$ up to a given order $\alpha_s \geq s$.
Looking at eqs.~(\ref{eq:00}) to (\ref{eq:1r}) we find the
following table:

\begin{center}
\begin{tabular}{|c|cc|}
\hline
$\alpha_s \geq s$ & $\omega^{i_{0}, \beta_0}$ & $\omega^{i_{1}, \beta_1}$ \\
\hline $d \omega^{k_{0}, \alpha_0}$ & $\beta_0 \leq \alpha_0$ &
$\beta_1 \leq \alpha_0$ \\
$d \omega^{k_{1}, \alpha_1}$ & $\beta_0 \leq \alpha_1-1$ &
$\beta_1 \leq \alpha_1$ \\
\hline
\end{tabular}
\\[6pt]
{\footnotesize Orders $\beta_p$ of the forms $\omega^{i_{p},
\beta_p}$ that contribute to $d\omega^{k_{s}, \alpha_s}$}
\end{center}

\noindent Since there must be enough one-forms in
(\ref{eq:largerbasis}) for the MC equations (\ref{eq:MCsub}) to be
satisfied, the $N_{0}+1$ and $N_{1}$ one-forms $\omega^{i_{0},
\alpha_0}$  ($\alpha_0=0, 1,\ldots,N_0$) and $\omega^{i_{1},
\alpha_1}$ ($\alpha_1=1,2,\ldots,N_1$) in (\ref{eq:largerbasis})
should include, at least, those appearing in their differentials.
Thus, the previous table implies the reverse inequalities
\begin{center}
\begin{tabular}{|c|cc|}
\hline
$\alpha_s \geq s$ & $\omega^{i_{0}, \beta_0}$ & $\omega^{i_{1}, \beta_1}$ \\
\hline
$d \omega^{k_{0}, \alpha_0}$ & $N_{0} \geq N_{0}$ & $N_{1} \geq N_{0}$ \\
$d \omega^{k_{1}, \alpha_1}$ & $N_{0}  \geq N_{1}-1$ & $N_{1} \geq N_{1}$ \\
\hline
\end{tabular}
\\[6pt]
{\footnotesize Conditions on the number $N_0$ $(N_1)$ of one-forms
$\omega^{i_{0}, \alpha_0} (\omega^{i_{1}, \alpha_1})$}
\end{center}
Hence, in this case there are two ways of cutting the expansions
(\ref{eq:serieszero}), (\ref{eq:seriesone}), namely for
{\setlength\arraycolsep{2pt}
\begin{eqnarray}
\label{eq:order1}
N_1 & = & N_0 \;, \\
\label{eq:order2} \textrm{or}\qquad N_1 & = & N_0 + 1 \; .
\end{eqnarray}}
\\[-8pt]
\noindent Besides (\ref{eq:N}) there is now an additional type of
solutions, eq.~(\ref{eq:order2}). For the $N_0=0, N_1=1$ values
eq.~(\ref{eq:dimsub}) yields $\textrm{dim} \, \mathcal{G}(0,1)=
\textrm{dim} \, \mathcal{G}$. Then, $\alpha_0=0$ and $\alpha_1=1$
only, the label $\alpha_p$ may be dropped and the structure
constants (\ref{eq:Csub}) for $\mathcal{G}(0,1)$ read
\begin{equation} \label{eq:CIW}
C_{i_{ p}\,j_{ q}}^{k_{ s}}= \left\{ \begin{array}{lll} 0, &
\mathrm{if} \ p + q \neq s  \\
c_{i_{ p}j_{ q}}^{k_{s}}, & \mathrm{if} \ p + q = s  \end{array}
\right. \quad \begin{array}{l}
p=0,1 \\
i_{p,q,s}=1,2,\ldots, \textrm{dim} \, V_{p,q,s} \quad ,
\end{array}
\end{equation}
which shows that $V_1$ is an abelian ideal of $\mathcal{G}(0,1)$.
Hence, $\mathcal{G}(0,1)$ is just the (simple) IW contraction of
$\mathcal{G}$ with respect to the subalgebra $\mathcal{L}_0$, as
it may be seen by taking the $\lambda\rightarrow 0$ limit in
(\ref{eq:MCins0})-(\ref{eq:MCins1}), which reduce to eqs.
(\ref{eq:00}) and (\ref{eq:11}). We can thus state the following
\begin{Th} \label{Th:contraction}
Let $\mathcal{G}=V_{0} \oplus V_{1}$, where $V_0$ is a subalgebra
$\mathcal{L}_{0}$. Let the coordinates $g^{i_p}$ of $G$ be
rescaled by $g^{i_0} \rightarrow  g^{i_{0}},g^{i_1} \rightarrow
\lambda g^{i_{1}}$ (eq.~(\ref{eq:redefinition})). Then, the
coefficient one-forms $\{\omega^{i_0,\alpha_0}$; $\omega^{i_1,
\alpha_1}\}$ of the expansions (\ref{eq:serieszero}),
(\ref{eq:seriesone}) of the Maurer-Cartan forms of $\mathcal{G}^*$
determine Lie algebras $\mathcal{G}(N_{0}, N_{1})$ when $N_1=N_0$
{\it or} $N_1=N_0+1$ of dimension $\textrm{dim} \,
\mathcal{G}(N_0,N_1)= (N_0+1) \, \textrm{dim} \, V_0 + N_1
\textrm{dim} \, V_1$ and with structure constants (\ref{eq:Csub}),
\begin{displaymath}
C_{i_{ p},\beta_p \; j_{ q},\gamma_q}^{k_{ s},\alpha_s}= \left\{
\begin{array}{lll} 0, &
\mathrm{if} \ \beta_p + \gamma_q \neq \alpha_s  \\
c_{i_{ p}j_{ q}}^{k_{s}}, & \mathrm{if} \ \beta_p + \gamma_q =
\alpha_s  \end{array} \right. \quad \begin{array}{l}
p,q,s=0,1 \\
i_{p,q,s}=1,2,\ldots, \textrm{dim} \, V_{p,q,s} \\
\alpha_0,\beta_0,\gamma_0=0,1, \ldots, N_0 \\
\alpha_1,\beta_1,\gamma_1=1,2, \ldots, N_1 \;  .
\end{array}
\end{displaymath}
       In particular, $\mathcal{G}(0,0)=\mathcal{L}_0$ and
$\mathcal{G}(0,1)$ (eq.~(\ref{eq:order2}) for $N_{0} =0$) is the
simple IW contraction of $\mathcal{G}$ with respect to the
subalgebra $\mathcal{L}_{0}$.
\end{Th}

\subsection{The case in which $V_1$ is a symmetric coset \label{scoset}}
Let us now  particularize to the case in which $\mathcal{G}/
\mathcal{L}_0=V_1$ is a symmetric coset {\it i.e.},
\begin{equation} \label{eq:z2grad2}
[V_{ 0}, \, V_{ 0}] \subset V_{ 0} \; , \quad [V_{ 0}, \, V_{ 1}]
\subset V_{ 1} \; , \quad [V_{ 1}, \, V_{ 1}] \subset V_{ 0} \; ,
       \end{equation}
$\left( [V_{ p}, \, V_{ q}] \subset V_{ p +  q} \; , (p+q)
\textrm{mod}\, 2 \right)$. This applies, for instance, to all
superalgebras where $V_0$ is the bosonic subspace and $V_1$ the
fermionic one. Then, if $c_{i_{ p}j_{ q}}^{k_{ s}}$ ($ p,q,s
=0,1$; $i_p=1,\ldots\textrm{dim} \, V_p$) are the structure
constants of $\mathcal{G}$, $c_{i_{ p}j_{q}}^{k_{ s}}=0$ if $ s
\neq (p+q) \textrm{mod} \, 2$, the MC equations reduce to
{\setlength\arraycolsep{2pt}
\begin{eqnarray}
\label{eq:Z2MCzero} d\omega^{k_{0}} & = & -\frac{1}{2}
c_{i_{0}j_{0}}^{k_{0}} \omega^{i_{0}} \wedge \omega^{j_{0}}
-\frac{1}{2} c_{i_{1}j_{1}}^{k_{0}}
\omega^{k_{1}} \wedge \omega^{j_{1}}  \\
\label{eq:Z2MCone} d\omega^{k_{1}} & = & - c_{i_{0}j_{1}}^{k_{1}}
\omega^{i_{0}} \wedge \omega^{j_{1}} \; ,
\end{eqnarray}}
\\[-8pt]
\noindent and one can state the following

\begin{prop} \label{prop:z2serie}
Let $G$ and $\mathcal{G}$ be as in Th.~\ref{Th:contraction}, and
let further $V_1$ be a symmetric space, eq.~(\ref{eq:z2grad2}).
Then, the rescaling (\ref{eq:redefinition}) leads to an even (odd)
power series in $\lambda$ for the MC forms
$\omega^{i_0}(g,\lambda)$ ($\omega^{i_1}(g,\lambda)$):
\begin{eqnarray} \label{eq:z2splitseries1}
\omega^{i_{ 0}}(g,\lambda) & = & \omega^{i_{ 0},0}(g) + \lambda^2
\omega^{i_{ 0},2}(g) + \lambda^4 \omega^{i_{ 0},4}(g)
+ \ldots \nonumber \\
\omega^{i_{ 1}}(g,\lambda) & = & \lambda \omega^{i_{ 1},1}(g) +
\lambda^3 \omega^{i_{ 1},3}(g) + \lambda^5 \omega^{i_{ 1},5}(g) +
\ldots \; ,
\end{eqnarray}

\noindent namely, $\omega^{i_{ \overline{\alpha}}}(g, \lambda) =
\sum_{\alpha=0}^{\infty} \lambda^{\alpha}
\omega^{i_{\overline{\alpha}},\alpha}(g)\,;\, \overline{ \alpha}=
\alpha \, (\mathrm{mod} \, 2)$.
\end{prop}

\noindent {\it Proof.}

\noindent Under (\ref{eq:redefinition}) $ dg^{i_{ 0}} \rightarrow
dg^{i_{ 0}} \, , \; dg^{i_{ 1}} \rightarrow \lambda \, dg^{i_{
1}}$, which contributes with $\lambda^0$ ($\lambda$) to
$\omega^{i_{ 0}}(g,\lambda)$ ($\omega^{i_{ 1}}(g,\lambda)$);
$c_{j_{ q} k_{ s}}^{i_{ p}}$ vanish trivially unless $p=(q+s)\,
\textrm{mod} \, 2$ . Then, under (\ref{eq:redefinition}), the
$g^{k_{ s}} dg^{j_{ q}}$ terms in (\ref{eq:serie2}) with one
$g^{k_s}$ rescale as
\begin{eqnarray}
       p = 0 \; & : &
c_{j_{ 0} k_{{0}}}^{i_{ 0}} \, g^{k_{ 0}} dg^{j_{ 0}} \rightarrow
c_{j_{ 0} k_{{0}}}^{i_{ 0}} \, g^{k_{ 0}} dg^{j_{ 0}} \nonumber \;
, \quad c_{j_{ 1} k_{{1}}}^{i_{ 0}} \, g^{k_{ 1}} dg^{j_{ 1}}
\rightarrow \lambda^2 c_{j_{ 1} k_{{1}}}^{i_{ 0}} \,
g^{k_{{1}}} dg^{j_{ 1}} \; ;\nonumber \\
       p =  1 \; & : &
c_{j_{ 0} k_{{1}}}^{i_{ 1}} \, g^{k_{ 1}} dg^{j_{ 0}}  \rightarrow
\lambda \,  c_{j_{ 0} k_{{1}}}^{i_{ 1}} \, g^{k_{{1}}} dg^{j_{ 0}}
\; ,
\end{eqnarray}

\noindent so that the  powers $\lambda^0$ and $\lambda^2$
($\lambda$) contribute to $\omega^{i_{ 0}}(g,\lambda)$
($\omega^{i_{ 1}}(g,\lambda)$). For the terms in (\ref{eq:serie2})
involving the products of $n$ $g^{k_s}$'s,

\begin{equation}
c_{j_{ q} k_{{s}_1}}^{h_{{t}_1}} c_{h_{{t}_1}
k_{{s}_2}}^{h_{{t}_2}} \ldots
        c_{h_{{t}_{n-2}} k_{{t}_{n-1}}}^{h_{{t}_{n-1}}}
c_{h_{{t}_{n-1}} k_{{s}_n}}^{i_{ p}} g^{k_{{s}_1}} g^{k_{{s}_2}}
\ldots g^{k_{{s}_{n-1}}}
        g^{k_{{s}_n}} dg^{j_{ q}} \; ,
\end{equation}
\noindent the fact that $V_1=\mathcal{G}/\mathcal{L}_0$ is a
symmetric space requires that $p=q + {s}_1 + {s}_2 \ldots + {s}_n
\, \mathrm{(mod 2)}$. Thus, after the rescaling
(\ref{eq:redefinition}), only even (odd) powers of $\lambda$, from
$\lambda^0$ ($\lambda$) up to the closest (lower or equal to)
$n+1$ even (odd) power $\lambda^{n+1}$, contribute to $\omega^{i_{
0}}(g,\lambda)$ ($\omega^{i_{ 1}}(g,\lambda)$), {\it q.e.d}.

\subsubsection{Structure of ${\mathcal G}(N_0,N_1)$ in the symmetric
coset case}

Inserting the power series above into the MC equations
(\ref{eq:Z2MCzero}) and (\ref{eq:Z2MCone}), we arrive at the
equalities: {\setlength\arraycolsep{0pt}
\begin{eqnarray}
\label{eq:z2even} &d&\omega^{k_{0}, 2\sigma} = -\frac{1}{2}
c_{i_{0}j_{0}}^{k_{0}} \sum_{\rho=0}^{\sigma} \omega^{i_{0},2
\rho} \wedge \omega^{j_{0},2(\sigma-\rho)} -\frac{1}{2}
c_{i_{1}j_{1}}^{k_{0}}
       \sum_{\rho=1}^{\sigma} \omega^{i_{1},2\rho-1} \wedge
\omega^{j_{1}, 2(\sigma-\rho)+1} \; ,  \\
\label{eq:z2odd} &d&\omega^{k_{1}, 2\sigma+1} =
-c_{i_{0}j_{1}}^{k_{1}} \sum_{\rho=0}^{\sigma} \omega^{i_{0},
2\rho} \wedge \omega^{j_{1},2(\sigma-\rho)+1} \; ,
\end{eqnarray}}

\noindent where the expansion orders $\alpha$ are either
$\alpha=2\sigma$ or $\alpha=2\sigma+1$. From them it follows that
the vector spaces generated by
\begin{equation} \label{eq:z2largerbasis}
\{ \omega^{i_{0}, 0}, \omega^{i_{0}, 2}, \omega^{i_{0}, 4},
\ldots, \omega^{i_{0}, N_{0}}; \omega^{i_{1}, 1}, \omega^{i_{1},
3}, \ldots, \omega^{i_{1}, N_{1}} \} \; ,
\end{equation}
\noindent where $N_0 \geq 0$ (and even) and $N_1 \geq 1$ (and
odd), will determine a Lie algebra when
\begin{eqnarray}
\label{eq:Nzeroz2}
N_1&=&N_0-1 \quad ,\\
\label{eq:Nonez2} \textrm{or} \quad N_1&=&N_0+1 \quad .
\end{eqnarray}

\noindent Notice that we have a new type of  solutions
(\ref{eq:Nzeroz2}) with respect to the preceding case
(eqs.~(\ref{eq:order1}), (\ref{eq:order2})), and that the previous
solution $N_0=N_1$ is not allowed now since $N_0$ ($N_1$) is
necessarily even (odd). Then, for the symmetric case, the algebras
${\mathcal G}(N_0,N_1)$ may also be denoted $\mathcal{G}(N)$,
where $N=\textrm{max} \{N_0,N_1 \}$, and are obtained at each
order by adding alternatively copies of $V_0$ and $V_1$. Its
structure constants are given by
\begin{equation}
C_{i_{\overline{\beta}},\beta \;
j_{\overline{\gamma}},\gamma}^{k_{\overline{\alpha}},\alpha}=
\left\{
\begin{array}{lll}
0,                        & \mathrm{if} \ \beta + \gamma \neq \alpha  \\
c_{i_{\overline{\beta}}j_{\overline{\gamma}}}^{k_{\overline{\alpha}}},
& \mathrm{if} \ \beta + \gamma = \alpha \; ; \;
\overline{\alpha}=\alpha\, (\textrm{mod}2),\,
\overline{\beta}=\beta \, (\textrm{mod}2),
\,\overline{\gamma}=\gamma\, (\textrm{mod}2) \end{array} \right.
\end{equation}

Let us write explicitly the MC eqs.~for the first algebras
obtained. If we allow for $N_1=0$, we get the trivial case

$\mathcal{G}(0,0)=\mathcal{G}(0)$:
\begin{equation} \label{eq:z2Nzero}
d\omega^{k_{0},0}= -\frac{1}{2} c_{i_{ 0}j_{ 0}}^{k_{ 0}}
\omega^{i_{ 0}, 0} \wedge \omega^{j_{ 0}, 0}
\end{equation}
{\it i.e.}, $\mathcal{G}(0,0)$ is the subalgebra $\mathcal{L}_0$
of the original algebra $\mathcal{G}$.

$\mathcal{G}(0,1)=\mathcal{G}(1)$: {\setlength\arraycolsep{2pt}
\begin{eqnarray} \label{eq:z2N10}
d\omega^{k_{0},0}&=& -\frac{1}{2} c_{i_{ 0}j_{ 0}}^{k_{ 0}}
\omega^{i_{ 0}, 0} \wedge \omega^{j_{
0}, 0} \; , \\
\label{eq:z2N11} d\omega^{k_{ 1}, 1}&=& -c_{i_{ 0}j_{ 1}}^{k_{ 1}}
\omega^{i_{ 0}, 0} \wedge \omega^{j_{ 1}, 1}\; ,
\end{eqnarray}}
\\[-8pt]
\noindent so that $\mathcal{G}(0,1)$ is again
(Th.~\ref{Th:contraction}) the IW contraction of $\mathcal{G}$
with respect to $\mathcal{L}_{0}$.

$\mathcal{G}(2,1)=\mathcal{G}(2)$: {\setlength\arraycolsep{0.5pt}
\begin{eqnarray} \label{eq:z2N20}
d\omega^{k_{ 0}, 0}&=& -\frac{1}{2}  c_{i_{ 0}j_{ 0}}^{k_{
0}} \omega^{i_{ 0}, 0} \wedge \omega^{j_{ 0}, 0} \; , \\
\label{eq:z2N21} d\omega^{k_{ 1}, 1}&=& -c_{i_{ 0}j_{ 1}}^{k_{1}}
\omega^{i_{ 0}, 0} \wedge \omega^{j_{ 1}, 1}\; ,\\
\label{eq:z2N22} d\omega^{k_{ 0}, 2}&=& -c_{i_{ 0}j_{ 0}}^{k_{ 0}}
\omega^{i_{ 0}, 0} \wedge \omega^{j_{ 0}, 2} -\frac{1}{2} c_{i_{
1}j_{1}}^{k_{ 0}} \omega^{i_{ 1}, 1} \wedge \omega^{j_{ 1}, 1}\; .
\end{eqnarray}}

The structure of the Lie algebras $\mathcal{G}(N)$ is given by the
following

\begin{prop}
The Lie algebra $\mathcal{G}(0)=\mathcal{L}_0$ is a subalgebra of
$\mathcal{G}(N)$ for all $N \ge 0$. $W_{\alpha}$ in (\ref{def:W})
reduces here to
\begin{equation}
W_{\alpha}=\left\{ \begin{array}{lll} V_{0,\alpha}, &
\mathrm{if} \ \alpha \, \mathrm{even} \\
V_{1,\alpha}, & \mathrm{if} \ \alpha \; \mathrm{odd} \quad .
\end{array} \right.
\end{equation}
For $N \ge 1$, $W_N$  is an abelian ideal $\mathcal{W}_N$ of
$\mathcal{G}(N)$ and $\mathcal{G}(N) / \mathcal{W}_N =
\mathcal{G}(N-1)$, {\it i.e.}, $\mathcal{G}(N)$ is an extension of
$\mathcal{G}(N-1)$ by $\mathcal{W}_N$.

Further, for $N$ even and $\mathcal{L}_{ 0}$ abelian, the
extension $\mathcal{G}(N)$ of $\mathcal{G}(N-1)$ by
$\mathcal{W}_N$ is central.
\end{prop}

\noindent \textit{Proof.}

\noindent The proof of the first part proceeds as in
Prop.~\ref{prop:ext}. For the second part, notice that, for $N
\geq 1$, the only thing that prevents the abelian ideal
$\mathcal{W}_N$ from being central is its failure to commute with
$\mathcal{W}_0 \approx \mathcal{L}_{ 0}$, since $[W_{\alpha}, \,
\mathcal{W}_N] =0$ for $\alpha=1,2,\ldots, N$. But for $N$ even,
$C_{i_0,0\;j_0,N}^{k_0,N}= c_{i_{0} j_{ 0}}^{k_{ 0}}$ , which
vanish for $\mathcal{L}_0$ abelian. Thus $\mathcal{W}_N$ becomes a
central ideal, and $\mathcal{G}(N)$ a central extension of
$\mathcal{G}(N-1)$ by $\mathcal{W}_N$, {\it q.e.d}.

\section{Different powers rescaling subordinated to a general splitting
of ${\mathcal G}$} \label{general}

Let us extend now the above results to the case where the group
parameters are multiplied by arbitrary integer powers of
$\lambda$. Let $\mathcal{G}$ be split into a sum of $n+1$ vector
subspaces,
\begin{equation} \label{eq:splitn}
{\mathcal G}=V_{0} \oplus V_{1} \oplus  \cdots \oplus V_{n}
=\bigoplus_0^n V_p \; ,
\end{equation}
and let the rescaling
\begin{equation} \label{eq:nredef}
g^{i_{ 0}} \rightarrow  g^{i_{ 0}} \; , \; g^{i_{ 1}} \rightarrow
\lambda g^{i_{ 1}} \; , \; \ldots , g^{i_{ n}} \rightarrow
\lambda^n g^{i_{ n}} \quad \quad (g^{i_{ p}} \rightarrow \lambda^p
g^{i_{ p}}, \; p=0,\ldots,n)
\end{equation}
    of the group coordinates $g^{i_{ p}}$ be
subordinated to the splitting (\ref{eq:splitn}) in an obvious way.
We found in the previous section ($p=0,1$) that, when the
rescaling (\ref{eq:redefinition}) was performed,  having $V_0$ as
a subalgebra $\mathcal{L}_0$ proved to be convenient (though not
necessary) since it led to more types of solutions
((\ref{eq:order1})-(\ref{eq:order2}),{\it cf.}~(\ref{eq:N})).
Furthermore, the first order algebra $\mathcal{G}(0,1)$ for that
case was found to be the simple IW contraction of $\mathcal{G}$
with respect to $\mathcal{L}_0$. By the same reason, we will
consider here conditions on $\mathcal{G}$ that will lead to a
richer new algebras structure, including the generalized IW
contraction of $\mathcal{G}$ in the sense \cite{Wei:00} of
Weimar-Woods (W-W)\footnote{This is, in fact, the most general
contraction: any contraction is equivalent to a generalized IW
contraction with integer exponents \cite{Wei:00}. The generalized
IW contraction is defined as follows. Let $\mathcal{G}=\oplus
V_p$, $p=0,1,\ldots,n$. Let the basis generators $X$ of
$\mathcal{G}$ of each subspace $V_p$ be redefined by $X
\rightarrow \lambda^{n_p}X$, where the $n_p$ may be chosen to be
integers. Then, it is evident that the generalized IW 
contraction (the limit $\lambda \rightarrow 0$) exists iff 
${\mathcal G}$ is such that $[V_p,V_q]\subset \oplus_s V_s$, 
where $s$ runs over all the values for which $n_s \leq n_p + n_q$. 
For (\ref{eq:splitn}), (\ref{eq:nredef}) and $n_p\equiv p$, this 
is equivalent to (\ref{eq:cont}) above.}. In terms of the 
structure constants of $\mathcal G$ we will then require
\begin{equation} \label{eq:cont}
c_{i_p j_q}^{k_s}=0 \quad \textrm{if $s > p+q$} \;
\end{equation}
{\it i.e.}, that the Lie bracket of elements in $V_p$, $V_q$ is in
$\oplus_s V_s$ for $s \leq p+q$. This condition leads, through
(\ref{eq:serie2}), to a power series expansion of the one-forms
$\omega^{i_{ p}}$ in $V^*_{ p}$ that, for each $p=0,1,\ldots,n$,
starts precisely with the power $\lambda^p$,
{\setlength\arraycolsep{2pt}
\begin{eqnarray} \label{eq:szero}
\omega^{i_{ 0}}(g,\lambda) & = & \sum_{\alpha=0}^{\infty}
\lambda^\alpha \omega^{i_{ 0},\alpha}(g) = \omega^{i_{ 0},0}(g) +
\lambda \omega^{i_{ 0},1}(g) + \lambda^2 \omega^{i_{ 0},2}(g) +
\ldots \; , \\ \label{eq:sone} \omega^{i_{ 1}}(g,\lambda) & = &
\sum_{\alpha=1}^{\infty} \lambda^\alpha \omega^{i_{ 1},\alpha}(g)
= \lambda \omega^{i_{ 1},1}(g) + \lambda^2 \omega^{i_{ 1},2}(g) +
\lambda^3 \omega^{i_{ 1},3}(g) + \ldots \; , \\
\cdots & & \nonumber \\
\label{eq:sn} \omega^{i_{ n}}(g,\lambda) & = &
\sum_{\alpha=n}^{\infty} \lambda^\alpha \omega^{i_{ n},\alpha}(g)
= \lambda^n \omega^{i_{ n},n}(g) + \lambda^{n+1} \omega^{i_{
n},n+1}(g) + \ldots \quad .
\end{eqnarray}}

We may extend all the sums so that they begin at $\alpha=0$ by
setting $ \omega^{i_p,\alpha}\equiv 0$ when $\alpha<p$. Then,
inserting the expansions of $\omega^{i_p,\alpha}$ in the MC
eqs.~and using (\ref{eq:sumatorio}) we get (\ref{eq:MCG}) for
$p,q,s=0,1,\ldots,n$. If we now introduce the notation
$\omega^{i_p,\alpha_p}$ with different ranges for the expansion
orders, $\alpha_p=p,p+1,\ldots N_p$ for each $p$, we see that the
MC eqs. take the form

\begin{equation} \label{eq:MCn}
d\omega^{k_{ s}, \alpha_s}= -\frac{1}{2} C_{i_{ p},\beta_p \;
j_{q},\gamma_q}^{k_{ s},\alpha_s}\; \omega^{i_{ p}, \beta_p}
\wedge \omega^{j_{ q}, \gamma_q} \quad ,
\end{equation}
where
\begin{equation} \label{eq:Cn}
C_{i_{ p},\beta_p \; j_{ q},\gamma_q}^{k_{ s},\alpha_s}= \left\{
\begin{array}{lll} 0, &
\mathrm{if} \ \beta_p + \gamma_q \neq \alpha_s  \\
c_{i_{ p}j_{ q}}^{k_{s}}, & \mathrm{if} \ \beta_p + \gamma_q =
\alpha_s  \end{array} \right. \quad \begin{array}{l}
p,q,s=0,1,\ldots, n \\
i_{p,q,s}=1,2,\ldots, \textrm{dim} \, V_{p,q,s} \\
\alpha_p,\beta_p,\gamma_p=p,p+1, \ldots, N_p
\end{array}
\end{equation}
and the $c_{i_p j_q}^{k_s}$ satisfy (\ref{eq:cont}). To find now
the $\omega^{i_p ,\beta_p}$'s that enter in $d
\omega^{k_s,\alpha_s}$, $s=0,1,\ldots,n$, we need an explicit
expression for it. This is found  in the Appendix, eqs.
(\ref{facil})-(\ref{eq:MCexp2}).
    From them we read that $d \omega^{k_{s}, \alpha_s}$,
$s=0,1,\ldots,n$, is expressed in terms of products of the forms
$\omega^{i_{p}, \beta_p}$ in the following table:

\begin{center}
\begin{tabular}{|c|ccccc|}
\hline
       $\alpha_s \geq s$ & $\omega^{i_{0}, \beta_0}$ & $\omega^{i_{1},
\beta_1}$
& $\omega^{i_{2}, \beta_2}$ & $\cdots$ & $\omega^{i_{n}, \beta_n}$  \\
\hline $d \omega^{k_{0}, \alpha_0}$ & $\beta_0 \leq \alpha_0$ &
$\beta_1 \leq \alpha_0$ &
$\beta_2 \leq \alpha_0$ & $\cdots$ & $\beta_n \leq \alpha_0$   \\
$d \omega^{k_{1}, \alpha_1}$ & $\beta_0 \leq \alpha_1-1$ &
$\beta_1 \leq \alpha_1$ & $\beta_2 \leq \alpha_1$ & $\cdots$
& $\beta_n \leq \alpha_1$ \\
$d \omega^{k_{2}, \alpha_2}$ & $\beta_0 \leq \alpha_2-2$ &
$\beta_1 \leq \alpha_2-1$ & $\beta_2 \leq \alpha_2$ & $\cdots$
& $\beta_n \leq \alpha_2$ \\
$\vdots$ & $\vdots$ & $\vdots$ & $\vdots$ & & $\vdots$  \\
$d \omega^{k_{n}, \alpha_n}$ & $\beta_0 \leq \alpha_n-n$ &
$\beta_1 \leq \alpha_n-n+1$ & $\beta_2
       \leq \alpha_n-n+2$ & $\cdots$ & $\beta_n \leq \alpha_n$ \\ \hline
\end{tabular}
\\[6pt]
{\footnotesize Types and orders of the forms $\omega^{i_{p},
\beta_p}$ needed to express $d\omega^{k_{s}, \alpha_s}$}
\end{center}

\noindent Now let $V^*(N_0, \ldots, N_n)$ be the vector space
generated by {\setlength\arraycolsep{2pt}
\begin{eqnarray} \label{eq:larger}
\{ \omega^{i_0 , \alpha_0} \ & ;& \omega^{i_1 , \alpha_1} \ ;
\ldots
; \omega^{i_n , \alpha_n} \}  =  \nonumber \\
& & \{ \omega^{i_{ 0}, 0}, \omega^{i_{ 0}, 1}, \stackrel{N_0
+1}{\ldots}, \omega^{i_{ 0}, N_{ 0}}; \, \omega^{i_{ 1}, 1},
\stackrel{N_1}{\ldots}, \omega^{i_{ 1}, N_{ 1}}; \, \ldots; \,
\omega^{i_{n}, n}, \stackrel{N_n-n+1}{\ldots}, \omega^{i_{n},
N_{n}} \} \; .
\end{eqnarray}}
These one-forms determine a Lie algebra
$\mathcal{G}(N_0,N_1,\ldots,N_n)$, of dimension
\begin{equation} \label{eq:dim}
\textrm{dim} \, \mathcal{G}(N_0, \ldots, N_n) = \sum_{p=0}^{n}
(N_p -p+1) \, \textrm{dim} \, V_p \;,
\end{equation}
under the conditions of the following

\begin{Th} \label{Th:newalg}
Let $\mathcal{G}=V_{0} \oplus V_{1} \oplus \cdots \oplus V_n$ be a
splitting of $\mathcal{G}$ into $n+1$ subspaces. Let $\mathcal{G}$
fulfill the Weimar-Woods contraction condition (\ref{eq:cont})
subordinated to this splitting, $c_{i_p j_q}^{k_s}=0$ if $s>p+q$.
The one-form coefficients $\omega^{i_p , \alpha_p}$ of
(\ref{eq:larger}) resulting from the expansion of the
Maurer-Cartan forms $\omega^{i_p}$ in which $g^{i_{ p}}
\rightarrow \lambda^p g^{i_{ p}}, \; p=0,\ldots,n$
(eq.~(\ref{eq:nredef})), determine Lie algebras
$\mathcal{G}(N_0,N_1,\ldots,N_n)$ of dimension (\ref{eq:dim}) and
structure constants
\begin{displaymath}
C_{i_{ p},\beta_p \; j_{ q},\gamma_q}^{k_{ s},\alpha_s}= \left\{
\begin{array}{lll} 0, &
\mathrm{if} \ \beta_p + \gamma_q \neq \alpha_s  \\
c_{i_{ p}j_{ q}}^{k_{s}}, & \mathrm{if} \ \beta_p + \gamma_q =
\alpha_s  \end{array} \right. \quad \begin{array}{l}
p,q,s=0,1,\ldots, n \\
i_{p,q,s}=1,2,\ldots, \textrm{dim} \, V_{p,q,s} \\
\alpha_p,\beta_p,\gamma_p=p,p+1, \ldots, N_p \quad ,
\end{array}
\end{displaymath}
(eq.(\ref{eq:Cn})) if $N_q=N_{q+1}$ {\it or} $N_q=N_{q+1}-1$
($q=0,1,\ldots,n-1$) in $(N_0,N_1,\ldots,N_n)$. In particular, the
$N_p=p$ solution determines the algebra $\mathcal{G}(0,1,\ldots,
n)$, which is the generalized \.In\"on\"u-Wigner contraction of
$\mathcal{G}$.
\end{Th}

\noindent {\it Proof}. \\
To enforce the closure under $d$ of the exterior algebra generated
by the one-forms in (\ref{eq:larger}) and to find the conditions
that the various $N_p$ must meet, we require, as in
Sec.~\ref{tres}, that all the forms $\omega^{i_p,\beta_p}$ present
in $d\omega^{k_s, \alpha_s}$ are already in (\ref{eq:larger}).
Looking at eqs.~(\ref{facil})-(\ref{eq:MCexp2}) and at the table
above, we find the restrictions

\begin{center}
\begin{tabular}{|c|ccccc|}
\hline
       $\alpha_s  \geq s$ & $\omega^{i_{0}, \beta_0}$ & $\omega^{i_{1},
\beta_1}$
& $\omega^{i_{2}, \beta_2}$ & $\cdots$ & $\omega^{i_{n}, \beta_n}$  \\
\hline $d \omega^{k_{0}, \alpha_0}$ & $N_0 \geq N_0$ & $N_1 \geq
N_0$ &
$N_2 \geq N_0$ & $\cdots$ & $N_n \geq N_0$   \\
$d \omega^{k_{1}, \alpha_1}$ & $N_0 \geq N_1 -1$ & $N_1 \geq N_1$
& $N_2 \geq N_1$
& $\cdots$ & $N_n \geq N_1$ \\
$d \omega^{k_{2}, \alpha_2}$ & $N_0 \geq N_2 -2$ & $N_1 \geq N_2
-1$ & $N_2 \geq N_2$
       & $\cdots$ & $N_n \geq N_2$ \\
$\vdots$ & $\vdots$ & $\vdots$ & $\vdots$ & & $\vdots$  \\
$d \omega^{k_{n}, \alpha_n}$ & $N_0 \geq N_n -n$ & $N_1 \geq N_n
-n+1$ & $N_2 \geq N_n -n+2$  & $\cdots$ & $N_n \geq N_n$ \\ \hline
\end{tabular}
\\[6pt]
{\footnotesize Closure conditions on the number $N_p$ of one-forms
$\omega^{i_{p}, \alpha_p}$}
\end{center}

\noindent It then follows that there are $2^n$ types of
solutions\footnote{\label{note} This number may be found, {\it
e.g.} for $n=3$, by writing symbolically the solution types in
(\ref{eq:gralcond}) as [0,0,0,0] for $N_0=N_1=N_2=N_3$; [0,0,0,1]
for $N_0=N_1=N_2, N_3=N_2+1$; [0,0,1,0] for $N_0=N_1,
N_2,=N_1+1=N_3$; [0,0,1,1] for $N_0=N_1, N_2,=N_1+1,N_3=N_2+1$;
[0,1,0,0] for $N_0,N_1=N_0+1=N_2=N_3$; [0,1,0,1] for
$N_0,N_1=N_0+1=N_2,N_3=N_2+1$; [0,1,1,0] for
$N_0,N_1=N_0+1,N_2=N_1+1=N_3$ and [0,1,1,1] for
$N_0,N_1=N_0+1,N_2=N_1+1,N_3=N_2+1$. This notation numbers the
solutions in base 2; since [0,1,1,1] corresponds to $2^3-1$ we
see, adding [0,0,0,0], that there are $2^3$ ways of cutting the
expansions that determine Lie algebras ${\mathcal
G}(N_0,N_1,N_2,N_3)$, and $2^n$ in the general ${\mathcal
G}(N_0,N_1,\ldots,N_n)$ case.} characterized by
$(N_0,N_1,\ldots,N_n)$, $N_p \geq p$, $p=0,1,\ldots,n,\,$ where

\begin{equation}
\label{eq:gralcond} N_{q+1} = N_q \quad \textrm{or} \quad N_{q+1}
= N_q +1 \quad (q=0,1,\ldots, n-1) \; .
\end{equation}
The JI for $\mathcal{G}(N_{0},\ldots, N_n)$,
{\setlength\arraycolsep{2pt}
\begin{eqnarray}\label{eq:granjacobi}
C_{i_p,\beta_p\;[j_q,\gamma_q}^{k_s,\alpha_s} C_{l_t,\rho_t\;
m_u,\sigma_u]}^{i_p,\beta_p} =0 &=& \nonumber \\
C_{i_p,\beta_p\;j_q,\gamma_q}^{k_s,\alpha_s} C_{l_t,\rho_t\;
m_u,\sigma_u}^{i_p,\beta_p} & + &
C_{i_p,\beta_p\;m_u,\sigma_u}^{k_s,\alpha_s}
C_{j_q,\gamma_q\;l_t,\rho_t}^{i_p,\beta_p}+
C_{i_p,\beta_p\;l_t,\rho_t}^{k_s,\alpha_s}
C_{m_u,\sigma_u\;j_q,\gamma_q}^{i_p,\beta_p} \; ,
\end{eqnarray}}
are again satisfied through the JI for $\mathcal{G}$. This is a
consequence of the fact that, for ${\mathcal G}$, the exterior
derivative of the $\lambda$-expansion of the MC eqs. is the
$\lambda$-expansion of their exterior derivative, but it may also
be seen directly\footnote{ We only need to check that
(\ref{eq:granjacobi}) reduces to the JI for $\mathcal{G}$ when the
order in the upper index is the sum of those in the lower ones
since the $C$'s are zero otherwise. First we see that, when
$\alpha_s=\gamma_q+\rho_t+\sigma_u$, all three terms in the {\it
r.h.s.} of (\ref{eq:granjacobi}) give non-zero contributions. This
is so because the range of $\beta_p$ is only limited by
$\beta_p\leq\alpha_s$, which holds when $\beta_p=\rho_t+\sigma_u$,
$\beta_p=\gamma_q+\rho_t$ and $\beta_p=\sigma_u+\gamma_q$.
Secondly, and since $\beta_p\geq p$, we also need that the terms
in the $i_p$ sum that are suppressed in (\ref{eq:granjacobi}) when
$p>\beta_p$ be also absent in the JI for $\mathcal{G}$ so that
(\ref{eq:granjacobi}) does reduce to the JI for $\mathcal{G}$.
Consider {\it e.g.}, the first term in the {\it r.h.s.}  of
(\ref{eq:granjacobi}). If $p>\beta_p$, then $p>\rho_t+\sigma_u$
and hence $p>t+u$. Thus, by the W-W condition (\ref{eq:cont}),
this term will not contribute to the JI for $\mathcal{G}$ and no
sum over the subspace $V_p$ index $i_p$ will be lost as a result.
The argument also applies to the other two terms for their
corresponding $\beta_p$'s.}.

A particular solution to (\ref{eq:gralcond}) is obtained by
setting $N_p = p$, $p=0,1,\ldots,n$, which defines
$\mathcal{G}(0,1,\ldots, n)$, with $\textrm{dim}
\,\mathcal{G}(0,1,\ldots, n)=\textrm{dim}\,\mathcal{G}=r$ (from
(\ref{eq:dim})). Since in this case $\alpha_p$ takes only one
value ($\alpha_p=N_p=p$) for each $p=0,1,\ldots,n$, we may drop
this label. Then, the structure constants (\ref{eq:Cn}) for
$\mathcal{G}(0,1, \ldots,n)$ read
\begin{equation} \label{eq:CIWn}
C_{i_{ p}\, j_{ q}}^{k_{ s}}= \left\{ \begin{array}{lll} 0, &
\mathrm{if} \ p + q \neq s  \\
c_{i_{ p}j_{ q}}^{k_{s}}, & \mathrm{if} \ p + q = s  \end{array}
\right. \quad \begin{array}{l}
p=0,1,\ldots,n \\
i_{p,q,s}=1,2,\ldots, \textrm{dim} \, V_{p,q,s} \; ,
\end{array}
\end{equation}
which shows that $\mathcal{G}(0,1,\ldots,n)$ is the generalized IW
contraction of $\mathcal{G}$, in the sense of \cite{Wei:00},
subordinated to the splitting (\ref{eq:splitn}). Of course, when
$n=1$ ($p=0,1$), $ V=V_0 \oplus V_1$,  $\mathcal{L}_0$ is a
subalgebra and eqs. (\ref{eq:gralcond}) ((\ref{eq:CIWn})) reduce
to (\ref{eq:order1}) or (\ref{eq:order2}) ((\ref{eq:CIW})), {\it
q.e.d.}

  Since the structure of $\mathcal{G}(N_0,N_1,\ldots,N_n)$ is fully
predetermined by  $\mathcal{G}$, we shall call 
$\mathcal{G}(N_0,N_1,\ldots,N_n)$ an {\it expansion} of $\mathcal{G}$. 
For instance, for the case ${\mathcal G}=V_0 \oplus V_1 \oplus V_2$ 
there are four types of expanded algebras 
${\mathcal G}(N_0,N_1,N_2)$\footnote{With the notation of footnote 
\ref{note}, these correspond, respectively, to [0,0,0], [0,0,1],[0,1,0] 
and [0,1,1].} 
{\setlength\arraycolsep{2pt}
\begin{eqnarray}
\label{eq:tres4}
N_0 & = & N_1 = N_2  \; \\
\label{eq:tres3}
N_0 & = & N_1 = N_2-1 \; ,  \\
\label{eq:tres2}
N_0 & = & N_1-1=N_2-1 \; ,  \\
\label{eq:tres1} N_0 & = & N_1-1= N_2-2 \; .
\end{eqnarray}}

Since in the above theorem $\alpha_p\geq p$ for all $p=0,\ldots,n$
was assumed, all types of one-forms $\omega^{i_p,\alpha_p}$ with
indexes $i_p$ in all subspaces $V_p$ were present in the basis of
$\mathcal{G}(N_0,N_1,\ldots,N_n)$. However, one may consider
keeping terms in the expansion up to a certain order $l,\; l<n$ in
which case due to (\ref{eq:sn}), the forms $\omega^{i_p,\alpha_p}$
with $p > l$ will not appear. Those with $p\leq l$ will determine
the vector space $V^*(N_0,N_1,\ldots,N_l)$ where $N_l$ is the
highest order $l$ and hence $\alpha_l$ takes only the value
$N_l=l=\alpha_l$. This vector space, of dimension
\begin{equation}\label{dimcajitas}
\textrm{dim}\,V^*(N_0, \ldots, N_l) = \sum_{p=0}^{l} (N_p -p+1) \,
\textrm{dim} \, V_p \quad,
\end{equation}
determines a Lie algebra $\mathcal{G}(N_0,N_1,\ldots,N_l)$ under
the conditions of the following theorem

\begin{Th}\label{Th:cajitas}
Let $\mathcal{G}=\oplus_0^n V_p$, etc.~as in Th.~\ref{Th:newalg}.
Then, up to a certain order $N_l=l<n$, the one-forms
\begin{equation} \label{lbasis4}
\{ \omega^{i_0 , \alpha_0} \ ; \omega^{i_1 , \alpha_1} \ ; \ldots
; \omega^{i_l , \alpha_l} \} = \{ \omega^{i_{ 0}, 0}, \omega^{i_{
0}, 1}, \stackrel{N_0 +1}{\ldots}, \omega^{i_{ 0}, N_{ 0}}\, ; \,
\omega^{i_{ 1}, 1}, \stackrel{N_1}{\ldots}, \omega^{i_{ 1}, N_{
1}}\, ; \, \ldots; \omega^{i_{l}, N_{l}} \} \; ,
\end{equation}
where $N_l=l=\alpha_l$, determine a Lie algebra
$\mathcal{G}(N_0,N_1,\ldots N_l)$ of dimension (\ref{dimcajitas})
and structure constants given by
\begin{equation} \label{eq:cnts4}
C_{i_{ p},\beta_p \; j_{ q},\gamma_q}^{k_{ s},\alpha_s}= \left\{
\begin{array}{lll} 0, &
\mathrm{if} \ \beta_p + \gamma_q \neq \alpha_s  \\
c_{i_{ p}j_{ q}}^{k_{s}}, & \mathrm{if} \ \beta_p + \gamma_q =
\alpha_s  \end{array} \right. \quad \begin{array}{l}
p,q,s=0,1,\ldots, l \\
i_{p,q,s}=1,2,\ldots, \textrm{dim} \, V_{p,q,s} \\
\alpha_p,\beta_p,\gamma_p=p,p+1, \ldots, N_p\,;\;N_p \le l \quad ,
\end{array}
\end{equation}
if $N_q=N_{q+1}$ or $N_q=N_{q+1}-1$, ($q=0,1,\ldots ,l-1$).
\end{Th}

\noindent {\it Proof}. \\
The restriction $\alpha_p \le N_l=l < n$ on the order $\alpha_p$
of the one-forms $\omega^{i_p,\alpha_p}$ implies, due to
(\ref{eq:sn}), that $V_l$ is monodimensional and that
$\omega^{i_l,l}$ is the last form entering (\ref{lbasis4}). Then,
looking at the closure conditions table in Th.~\ref{Th:newalg}, we
can restrict ourselves to the box delimited by
$\omega^{i_p,\beta_p}$, $d\omega^{k_s,\alpha_s}$ with $p,s \le l$.
This box will give spaces $V^*(N_0,N_1,\ldots, N_l)$, where
$N_q=N_{q+1}$ or $N_q=N_{q+1}-1$ ($q=0,1,\ldots,l-1$), and these
spaces will determine Lie algebras if the JI for (\ref{eq:cnts4})
\begin{equation} \label{JIcaja}
C_{i_p,\beta_p\; [ j_q,\gamma_q}^{k_s,\alpha_s} C_{l_t,\rho_t\;
m_u,\sigma_u ]}^{i_p,\beta_p} =0 \; ,\;\; i_{p,q,s}=1,2,\ldots,
\mathrm{dim} V_{p,q,s}
\end{equation}
{\it i.e.}, if $c_{i_p [ j_q}^{k_s} c_{l_t m_u]}^{i_p}=0\,, \;
s,q,t,u \le l$, is satisfied when
$\alpha_s=\gamma_q+\rho_t+\sigma_u$ above. Note that this is not
the JI for $\mathcal{G}$ since $i_p$ now runs over the basis of
$\oplus_0^l V_p \subset \mathcal{G}$ only since $p \le l$, and we
are thus removing the values corresponding to the basis of
$\oplus_{l+1}^n V_p$. However, if $p>l$ it is also {\it e.g.}
$p>\beta_p=\rho_t+\sigma_u \ge t+u$ in which case $c_{l_t\;
m_u}^{i_p}=0$ by (\ref{eq:cont}), {\it q.e.d.}

Since the structure constants (\ref{eq:cnts4}) are obtained from
those of $\mathcal{G}$ by restricting the $i_p$ indexes to be in
the subspaces $V_p,\, p\leq l$, $\mathcal{G}(N_0,N_1,\ldots,N_l)$
is {\it not} a subalgebra of $\mathcal{G}(N_0,N_1,\ldots,N_n)$.

\section{The expansion method for superalgebras}
\label{susy}

The above general procedure of generating Lie algebras from a
given one does not lie on the antisymmetry of the structure
constants of the original Lie algebra. Hence, with the appropriate
changes to account for the grading, the method is applicable when
$\mathcal{G}$ is a Lie superalgebra, a case which we consider in
this section.

Let $G$ be a supergroup and $\mathcal G$ its superalgebra. It is
natural to consider a splitting of $\mathcal G$ into the sum of
three subspaces $\mathcal G=V_0 \oplus V_1 \oplus V_2$, $V_1$
being the fermionic part of $\mathcal{G}$ and $V_0 \oplus V_2$ the
bosonic part, so that the notation reflects the
$\mathbb{Z}_2$-grading of $\mathcal{G}$. The even space is always
a subalgebra of $\mathcal{G}$ but it may be convenient to consider
it further split into the sum $V_0 \oplus V_2$ to allow for the
case in which a subspace ($V_0$) of the bosonic space is itself a
subalgebra $\mathcal{L}_0$.

Notice that, since $V_0$ is a Lie algebra $\mathcal{L}_0$, the
$\mathbb{Z}_2$-graduation of $\mathcal{G}$ implies that the
splitting $\mathcal G=V_0 \oplus V_1 \oplus V_2$ satisfies the W-W
contraction conditions (\ref{eq:cont}). Indeed, let $c_{i_p
j_q}^{k_s}$ ($i_{p,q,s}= 1, \ldots, \textrm{dim} \, V_{p,q,s}$,
$p,q,s=0,1,2$) be the structure constants of $\mathcal{G}$. The
$\mathbb{Z}_2$-graduation of $\mathcal G$ obviously  implies
{\setlength\arraycolsep{2pt}
\begin{eqnarray}
c_{i_0 j_0}^{k_1} &=& c_{i_0 j_1}^{k_2} =0 \label{eq:c1}\\
c_{i_0 j_1}^{k_0} &=&c_{i_1 j_1}^{k_1}= c_{i_0 j_2}^{k_1} =c_{i_2
j_1}^{k_0}= c_{i_2 j_1}^{k_2}= c_{i_2 j_2}^{k_1} =0
    \; .\label{eq:c2}
\end{eqnarray}

\noindent The first set of restrictions (\ref{eq:c1}), together
with the assumed subalgebra condition for $V_0$ (which, in
addition, requires $c_{i_0 j_0}^{k_2} =0$), are indeed the W-W
conditions (\ref{eq:cont}) for $\mathcal{G}$; note that these
conditions alone allow for $c_{i_1 j_1}^{k_0} \neq 0$, and
$c_{i_1j_1}^{k_2} \neq 0$ (and for $c_{i_1 j_1}^{k_1} \neq 0$,
although here $c_{i_1 j_1}^{k_1} = 0$ due to the
$\mathbb{Z}_2$-grading).

To apply now the above general procedure one must rescale the
group parameters. The rescaling (\ref{eq:nredef}) for $V=V_0
\oplus V_1 \oplus V_2$ takes the form
\begin{equation} \label{eq:superredef1}
g^{i_{ 0}} \rightarrow  g^{i_{ 0}} \; , \; g^{i_{ 1}} \rightarrow
\lambda g^{i_{ 1}} \; , \; g^{i_{ 2}} \rightarrow \lambda^2 g^{i_{
2}} \; .
\end{equation}
Note that, if it proves convenient on dimensional grounds (a
dimensionful parameter may be used to introduce dimensions in an
originally dimensionless algebra), the redefinitions
(\ref{eq:superredef1}) may be changed. They are equivalent {\it
e.g.}, to
\begin{equation} \label{eq:superredef2}
g^{i_{ 0}} \rightarrow  g^{i_{ 0}} \; , \; g^{i_{ 1}} \rightarrow
\mu^{1/2} g^{i_{ 1}} \; , \; g^{i_{ 2}} \rightarrow \mu g^{i_{ 2}}
\; ,
\end{equation}
with $ \mu=\lambda^2$ obviously suggested by the
$\mathbb{Z}_2$-graded commutators; other redefinitions are equally
possible.

The present $\mathbb{Z}_2$-graded case fits into the preceding
general discussion for $n=2$, but with additional restrictions
besides the W-W ones that follow from to the
$\mathbb{Z}_2$-grading.

\begin{Th}
\label{Th:super} Let $\mathcal{G}=V_{0}\oplus V_{1} \oplus V_2$ be
a Lie superalgebra, $V_{1}$ its odd part, and $V_{0} \oplus V_{2}$
the even one. Let further $V_0$ be a subalgebra $\mathcal{L}_0$.
As a result, $\mathcal{G}$ satisfies the W-W conditions
(\ref{eq:cont}) and, further, $V_1$ is a symmetric coset. Then,
the coefficients of the expansion of the Maurer-Cartan forms of
$\mathcal{G}$ rescaled by (\ref{eq:superredef1}) determine Lie
superalgebras $\mathcal{G}(N_{0},N_{1},N_2)$, $N_p\geq p,\,
p=0,1,2$, of dimension
\begin{equation}
\label{eq:dimsuper} \textrm{dim}\,\mathcal{G}(N_0,N_1,N_2)=
\left[\frac{N_0+2}{2}\right]\textrm{dim}V_0 +
\left[\frac{N_1+1}{2}\right]\textrm{dim}V_1 +
\left[\frac{N_2}{2}\right]\textrm{dim}V_2 \quad ,
\end{equation}
\noindent and structure constants
\begin{equation} \label{eq:Csuper}
C_{i_{ p},\beta_p \; j_{ q},\gamma_q}^{k_{ s},\alpha_s}= \left\{
\begin{array}{lll} 0, &
\mathrm{if} \ \beta_p + \gamma_q \neq \alpha_s  \\
c_{i_{ p}j_{ q}}^{k_{s}}, & \mathrm{if} \ \beta_p + \gamma_q =
\alpha_s  \end{array} \right. \quad \begin{array}{l}
p,q,s=0,1,2 \\
i_{p,q,s}=1,2,\ldots, \textrm{dim} \, V_{p,q,s} \\
\alpha_p,\beta_p,\gamma_p=p,p+2, \ldots, N_p-2, N_p \; ,
\end{array}
\end{equation}
where\footnote{For the rescaling (\ref{eq:superredef2}) the orders
would be $\alpha_p=\frac{p}{2},\frac{p+2}{2},\ldots,
\frac{N_p}{2}$.} [$\quad$] denotes integer part and the $N_0,N_2$
(even) and $N_1$ (odd) integers satisfy one of the three
conditions below {\setlength\arraycolsep{2pt}
\begin{eqnarray}
\label{eq:tress1}
N_0 & = & N_1+1 = N_2  \; \\
\label{eq:tress2}
N_0 & = & N_1-1 = N_2 \; ,  \\
\label{eq:tress3} N_0 & = & N_1-1= N_2-2 \; .
\end{eqnarray}}
\\[-14pt]
\noindent In particular, the superalgebra $\mathcal{G}(0,1,2)$
(eq.~(\ref{eq:tress3}) for $N_{0}=0$) is the generalized
\.In\"on\"u-Wigner contraction of $\mathcal{G}$.
\end{Th}

\noindent {\it Proof.} \noindent Since $V_1$ is a symmetric coset
the rescaling (\ref{eq:superredef1}) leads to an even (odd) power
series in $\lambda$ for the one-forms $\omega^{i_0}(g,\lambda)$
and $\omega^{i_2}(g,\lambda)$ ($\omega^{i_1}(g,\lambda)$), as in
Sec.~\ref{scoset} (eqs.~(\ref{eq:z2splitseries1})). Thus, the
conditions $N_0, N_2$ even, $N_1$ odd, have to be added to those
that follow from the closure inequalities table in
Th.~\ref{Th:newalg}. This gives the conditions
\begin{eqnarray}
N_0+1 &\ge& N_1 \ge N_0-1 \\
N_1+1 &\ge& N_2 \ge N_1-1 \\
N_0+2 &\ge& N_2 \ge N_0 \; ,
\end{eqnarray}
\noindent from which eqs. (\ref{eq:tress1})-(\ref{eq:tress3})
follow, {\it q.e.d}.

\section{Application: the M-theory superalgebra}
\label{seis}

Let us apply these ideas to the case of the M-theory superalgebra
(see \cite{To}\cite{AF82,vHvP82,Bars97,Se97}). String/M theory
implies that the $D$=11 supersymmetry algebra of nature may be one
that includes the super-Poincar\'e algebra plus some additional
`central' bosonic generators. This (528+32+55)-dimensional algebra
may be described by its MC equations
\begin{eqnarray}
d\Pi^{\alpha\beta} &=&
-\frac{1}{4}{(\gamma^{\mu\nu})^\alpha}_\gamma
\sigma_{\mu\nu}\wedge \Pi^{\gamma\beta}
-\frac{1}{4}{(\gamma^{\mu\nu})^\beta}_\gamma \sigma_{\mu\nu}\wedge
\Pi^{\gamma\alpha}-\pi^\alpha \wedge \pi^\beta\nonumber\\
d\pi^\alpha &=& -\frac{1}{4}{(\gamma^{\mu\nu})^\alpha}_\beta
\sigma_{\mu\nu}\wedge \pi^\beta\nonumber\\
d\sigma_{\mu\nu} &=& -{\sigma_\mu}^\rho \wedge \sigma_{\rho\nu}
\quad (\mu,\nu=0,\ldots,10 \;;\;\alpha,\beta=1,\dots,32)\quad ,
\label{Malg}
\end{eqnarray}
where $\Pi^{\alpha\beta}=\Pi^{\beta\alpha}$ is a set of bosonic MC
forms that may be expanded as\footnote{Then, we have
$\Pi_\mu=\Pi^{\alpha\beta}(\gamma_{\mu})_{\alpha\beta}$,
$\Pi_{\mu\nu}=\Pi^{\alpha\beta} (\gamma_{\mu\nu})_{\alpha\beta}$
and $\Pi_{\mu_1\dots\mu_5} =
\Pi^{\alpha\beta}(\gamma_{\mu_1\dots\mu_5})_{\alpha\beta}$; this
breaks the $GL(32,\mathbb{R})$ invariance of $\Pi^{\alpha\beta}$
down to $SO(1,10)$. For an analysis that uses the maximal
automorphism group $GL(32,\mathbb{R})$ of the M-algebra without
the Lorentz part, see \cite{BdeAIL01}.} \break
$\Pi^{\alpha\beta}=-\frac{1}{32}\left(\Pi_{\mu}\gamma^\mu-
\frac{1}{2}\Pi_{\mu\nu}\gamma^{\mu\nu} +
\frac{1}{5!}\Pi_{\mu_1\dots\mu_5}
\gamma^{\mu_1\dots\mu_5}\right)^{\alpha\beta}$, $\pi^\alpha$ is a
spinorial fermionic\footnote{The complex conjugation of a product
is defined here as the product of the complex conjugates, without
reversing the order.} MC form and $\sigma_{\mu\nu}$ are the MC
forms of the Lorentz generators.  The spinor indexes
$\alpha,\beta$ are raised and lowered with the $D=11$ $32\times
32$ skewsymmetric matrix $C_{\alpha\beta}$; the algebra
(\ref{Malg}) has dimension $560+55$. The M theory superalgebra is
sometimes regarded (see {\it e.g.} \cite{To}) as an IW contraction
of the superalgebra $osp(1|32)$, of dimension $560$, given by the
528+32 MC equations
\begin{eqnarray}
          d\rho^{\alpha\beta} &=&
-{\rho^\alpha}_\gamma \wedge \rho^{\gamma\beta}-\nu^\alpha \wedge
\nu^\beta \nonumber\\
     d\nu^\alpha &=& -{\rho^\alpha}_\beta\wedge \nu^\beta
     \qquad (\alpha,\beta=1,\ldots,32)\;,
\label{ospmaurer}
\end{eqnarray}
with $\rho^{\alpha\beta}=\rho^{\beta\alpha}$ \hfill bosonic \hfill
and $\nu^\alpha$ \hfill  fermionic \hfill  or, \hfill  using
\hfill
$\rho^{\alpha\beta}=-\frac{1}{32}\left(\rho_{\mu}\gamma^\mu-
\frac{1}{2}\rho_{\mu\nu}\gamma^{\mu\nu} + \right.$ $\left.
\frac{1}{5!}\rho_{\mu_1\dots\mu_5}
\gamma^{\mu_1\dots\mu_5}\right)^{\alpha\beta}$ and taking $\gamma$
matrices such that $\gamma^{\mu_1\dots\mu_{11}}=
\epsilon^{\mu_1\dots\mu_{11}}$,
\begin{eqnarray}   \label{ospmaurerd}
       d\rho_{\mu} &=& -\frac{1}{16} \rho_\nu\wedge
{\rho^\nu}_\mu +\frac{1}{32(5!)^2}
{\epsilon^{\mu_1\dots\mu_{10}}}_\mu \rho_{\mu_1\dots\mu_5} \wedge
\rho_{\mu_6\dots \mu_{10}} -\nu^\alpha
(\gamma_\mu)_{\alpha\beta}\wedge \nu^\beta \nonumber\\
     d\rho_{\mu\nu} &=& -\frac{1}{16} \rho_\mu\wedge \rho_\nu -
\frac{1}{16} \rho_{\mu\sigma}\wedge {\rho^\sigma}_{\nu} -
\frac{1}{16(4!)}\rho_{\mu\mu_1\dots \mu_4}\wedge {\rho^{\mu_4\dots
\mu_1}}_\nu- \nu^\alpha
(\gamma_{\mu\nu})_{\alpha\beta}\wedge \nu^\beta \nonumber\\
   d\rho_{\mu_1\dots\mu_5} &=&
\frac{1}{16(5!)}{\epsilon^{\sigma\nu_1\dots\nu_5}}_{\mu_1 \dots
\mu_5} \rho_\sigma \wedge \rho_{\nu_1\dots\nu_5} +
\frac{5}{16}{\rho^\nu}_{[\mu_1\dots \mu_4} \wedge
\rho_{\mu_5]\nu}\nonumber \\
   & &+\frac{1}{4(4!)^2}
{\epsilon^{\nu_1\dots\nu_6}}_{\mu_1\dots\mu_5}
\rho_{\nu_1\nu_2\nu_3\sigma_1\sigma_2}\wedge
{\rho^{\sigma_2\sigma_1}}_{\nu_4\nu_5\nu_6}- \nu^\alpha
(\gamma_{\mu_1\dots\mu_5})_{\alpha\beta}\wedge \nu^\beta
\nonumber\\
d\nu^\alpha &=& \frac{1}{32}{\left(\rho_{\mu}\gamma^\mu-
\frac{1}{2}\rho_{\mu\nu}\gamma^{\mu\nu} +
\frac{1}{5!}\rho_{\mu_1\dots\mu_5}
\gamma^{\mu_1\dots\mu_5}\right)^\alpha}_\beta \wedge \nu^\beta \ .
\end{eqnarray}
   But this is so
provided one {\it excludes} from (\ref{Malg}) the  55 Lorentz
generators $\sigma_{\mu\nu}$; otherwise, there are not enough
generators in $osp(1|32)$ to give the M-algebra (\ref{Malg}) by
contraction.

We show now, however, that the expansion method allows us to
obtain the M-theory superalgebra from $osp(1|32)$. Let us divide
the $osp(1|32)$ vector space into three subspaces $V_0$, $V_1$ and
$V_2$ as in Sec. \ref{susy}. Let then $V_0^*$ be the space
generated by the 55 MC forms $\rho^{\mu\nu}=\rho^{\alpha\beta}
(\gamma^{\mu\nu})_{\alpha\beta}$ of the Lorentz subalgebra of
$osp(1|32)$, $V_1^*$ the fermionic subspace generated by
$\nu^\alpha$,  and $V_2^*$ the space generated by the remaining
11+462 bosonic generators
$\rho^\mu=\rho^{\alpha\beta}(\gamma^\mu)_{\alpha\beta}$,
$\rho^{\mu_1\dots\mu_5}=\rho^{\alpha\beta}
(\gamma^{\mu_1\dots\mu_5})_{\alpha\beta}$. Since, on the other
hand, $V_1$ is a symmetric coset (Sec.~\ref{scoset}), it follows
that the expansions of the forms in $V_0^*$ contain even powers of
$\lambda$ starting from $\lambda^0$, that those of the forms in
$V_1^*$ include only odd powers in $\lambda$ starting from
$\lambda^1$, and that those of $V_2^*$ contain even orders
starting with $\lambda^2$, {\it i.e.}
\begin{eqnarray}
\label{eq:Vcero}
    V_0^* & : & \quad
    \rho^{\mu\nu} = \sum^\infty_{n=0} \lambda^{2n} \rho^{\mu\nu,2n}
       \quad ;  \\
\label{Vuno} V_1^* & : &  \quad \nu^\alpha = \sum^\infty_{n=0}
\lambda^{2n+1}\nu^{\alpha,2n+1}
    \quad ; \\
\label{eq:Vdos} V_2^* & : & \quad
      \rho^\mu = \sum^\infty_{n=1}\lambda^{2n}\rho^{\mu,2n} \quad,\quad
\rho^{\mu_1\dots\mu_5} =
\sum^\infty_{n=1}\lambda^{2n}\rho^{\mu_1\dots\mu_5,2n} \quad .
\end{eqnarray}
\noindent Setting now $\lambda=\mu^{1/2}$, one may rewrite the
series (\ref{eq:Vcero})-(\ref{eq:Vdos}) as
\begin{eqnarray}
     \rho^{\alpha\beta} &=&
\frac{1}{64}(\gamma_{\mu\nu})^{\alpha\beta} \rho^{\mu\nu,0}
+ \sum^\infty_{n=1}\mu^n\rho^{\alpha\beta,n} \label{rhoexp}\\
\nu^\alpha &=& \sum^\infty_{n=0} \mu^{n+\frac{1}{2}}
\nu^{\alpha,n+\frac{1}{2}} \ , \label{nuexp}
\end{eqnarray}

\noindent where $\rho^{\alpha\beta,n}$ for $n \ge 1$ collects the
contributions from both (\ref{eq:Vcero}), (\ref{eq:Vdos}). We see
now that, by Th.~\ref{Th:super}, we may keep powers up to $n=1$ in
(\ref{rhoexp}) and $n=0$ in (\ref{nuexp}), since this corresponds
to taking $N_0=N_2=2,\,N_1=1$ and, by eq.~(\ref{eq:dimsuper}),
$\textrm{dim}\,{\mathcal G}(2,1,2)$=110+32+473 =560+55. So
renaming $\nu^{\alpha,1/2}=\pi^\alpha$,
$\rho^{\mu\nu,0}=16\sigma^{\mu\nu}$, $\rho^{\alpha\beta,1}=
\Pi^{\alpha\beta}$ and using eqs. (\ref{eq:MCn}) and (\ref{eq:Cn})
we obtain the M-algebra (\ref{Malg}).

One may also consider higher orders in $\mu$. A result, omitting
the Lorentz part, was given in \cite{HS}. There, the expansion
(\ref{rhoexp})-(\ref{nuexp}) was considered up to the power
$\mu^{7/2}$ (without the term $n=0$ for $\rho^{\mu\nu}$ which gives
the Lorentz subalgebra), in an attempt to derive the M-algebra of
Sezgin \cite{Se97}. This superalgebra (not to be confused with the
much smaller M-theory superalgebra of eq. (\ref{Malg})) was
constructed generalizing earlier results in \cite{AT89,BESE} in
order to re-interpret the FDA's associated to the WZ forms of
supersymmetric extended objects and supergravity as Lie algebras
(see also \cite{CAIPB00},\cite{AF82}). The superalgebra obtained
in \cite{HS} is not that in \cite{Se97}, but a subalgebra of it.
Sezgin's superalgebras, and their associated enlarged superspace
groups, are obtained by the third procedure in the Introduction:
they are superalgebra and supergroup extensions \cite{CAIPB00}.

    It may be shown that, in contrast with Sezgin's M-algebra,
the superalgebra in \cite{HS} is not enough to write the three-
and six-forms of the $D=11$ membrane and five-brane in flat
superspace in terms of invariant one-forms. In fact, the extended
algebras considered in \cite{BESE,CAIPB00} cannot be obtained by
the expansion method but for some exceptions (for example, it may
be seen that the Green algebra \cite{GRE} in three spacetime
dimensions may be obtained  from $osp(1|2)$). Nevertheless, we
have seen here that the Lorentz part of the M-theory superalgebra,
which is missing in the IW contraction of $osp(1|32)$, may be
generated when suitable orders of the series expansions of its MC
forms are retained: the M-algebra (\ref{Malg}) is
$osp(1|32)(2,1,2)$.

To conclude, we mention that another M-type algebra having the
528-dimensional $sp(32)$ as its automorphism group, namely
\begin{eqnarray} \label{spM}
d\Pi^{\alpha\beta} &=& -{\sigma^\alpha}_\gamma \wedge
\Pi^{\gamma\beta} -{\sigma^\beta}_\gamma \wedge \Pi^{\gamma\alpha}
-\pi^\alpha\wedge\pi^\beta \nonumber\\
d\pi^\alpha &=& - {\sigma^\alpha}_\gamma \wedge \pi^\gamma   \nonumber\\
d\sigma^{\alpha\beta} &=& - {\sigma^\alpha}_\gamma \wedge
\sigma^{\gamma\beta} \quad ,
\end{eqnarray}\\[-20pt]
\noindent may also be obtained using the natural splitting of
$osp(1|32)$ in its even and odd parts. In the notation of
Sec.~\ref{scoset} (expansion in $\lambda$) the superalgebra
(\ref{spM}) is $osp(1|32)$(2,1), where now
$\rho^{\alpha\beta,0}=\sigma^{\alpha\beta}$ corresponds to the
$sp(32)$ subalgebra, $\rho^{\alpha\beta,2}=\Pi^{\alpha\beta}$ and
$\nu^{\alpha,1}=\pi^{\alpha}$.

\section{Extension to gauge free differential (super)algebras
and Chern-Simons theories} \label{gfda}

We have shown that, by rescaling some group variables and
identifying equal powers of $\lambda$ in the MC equations of an
algebra $\cal G$, one obtains the MC equations (\ref{eq:MCn}) for
the algebra $\mathcal{G}(N_0,N_1,\dots, N_n)$. Using these MC
equations we may now construct the corresponding gauge free
differential (super)algebras (for FDA see
\cite{Su77,AF82,Ni83,Cd'AF91}).

Let us examine the general case of Sec.~4. To obtain a gauge FDA
we replace the MC forms $\omega^{k_s,\alpha_s}$ by the gauge field
(or `soft', see \cite{Cd'AF91}) one-forms $A^{k_s,\alpha_s}$ and
introduce their corresponding curvatures $F^{k_s,\alpha_s}$ by
({\it cf.} (\ref{eq:MCn}))
\begin{equation}
\label{curvatures} F^{k_{ s}, \alpha_s} =dA^{k_{ s}, \alpha_s}+
\frac{1}{2} C_{i_{ p},\beta_p \; j_{ q},\gamma_q}^{k_{
s},\alpha_s}\; A^{i_{ p}, \beta_p} \wedge A^{j_{ q},\gamma_q}
\,:=\, DA^{k_{ s}, \alpha_s} \quad,
\end{equation}
where the $C_{i_{ p},\beta_p \; j_{ q},\gamma_q}^{k_{
s},\alpha_s}$ are defined as in (\ref{eq:Cn}). The curvatures
$F^{k_s,\alpha_s}$ satisfy the consistency conditions expressed by
the Bianchi identities

\begin{equation}
    \label{Bianchi}
      dF^{k_{ s}, \alpha_s} =
    C_{i_{ p},\beta_p\; j_{ q},\gamma_q}^{k_{ s},\alpha_s}\;
      F^{i_{ p}, \beta_p} \wedge A^{j_{ q}, \gamma_q}\quad, \quad (DF=0)
\quad.
\end{equation}
The Cartan structure equations (\ref{curvatures}) and the Bianchi
identities (\ref{Bianchi}) define the gauge FDA associated with
${\mathcal G}(N_0,N_1,\ldots,N_n)$.

   One may look at eqs. (\ref{curvatures})-(\ref{Bianchi})
as coming from expansions of the type
(\ref{eq:szero})-(\ref{eq:sn}) of the original gauge FDA for
${\mathcal G}$,
\begin{equation}
\label{FDAoriginal} F^{k_s} =  dA^{k_s}+\frac{1}{2} c_{i_p\;
j_q}^{k_s}\; A^{i_p} \wedge A^{j_q} \quad,\quad
     dF^{k_s} =
    c_{i_p\; j_q}^{k_s}\; F^{i_p} \wedge A^{j_q}  \quad ,
\end{equation}
\noindent or directly as defining the gauge FDA associated with
the ${\mathcal G}(N_0,N_1,\ldots,N_n)$ Lie algebra. Moreover, the
infinitesimal gauge transformations of the $A^{k_s,\alpha_s}$,
$F^{k_s,\alpha_s}$, with parameters $\varphi^{k_s,\alpha_s}$
corresponding to the expanded group $G(N_0,N_1\ldots,N_n)$, are 
given by
\begin{equation}
\delta A^{k_s, \alpha_s} = d\varphi^{k_s, \alpha_s}
    -C_{i_p,\beta_p\; j_q,\gamma_q}^{k_s,\alpha_s}\;
\varphi^{i_p, \beta_p} A^{j_{ q}, \gamma_q}\; ,\;
\delta F^{k_{ s}, \alpha_s} =
C_{i_{ p},\beta_p\; j_{ q},\gamma_q}^{k_{ s},\alpha_s}\;
F^{i_{ p}, \beta_p} \varphi^{j_{ q}, \gamma_q} \;,
\label{transf}
\end{equation}
recalling that the original gauge fields $A^{k_s}$ of $G$ transform as
\begin{equation}
\delta A^{k_s} = d\varphi^{k_s}-c^{k_s}_{i_p j_q}\varphi^{i_p}
A^{j_q} \ . \label{transfg}
\end{equation}
\noindent We note that the above FDA algebras are {\it
contractible} \cite{Su77,Ni83,Cd'AF91} since they are generated by
pairs of forms $(A,B)$ such that $B=dA$ and $dB=0$, where $A$
corresponds to $A^{k_s,\alpha_s}$ and $B=dA=F-A^2$ ({\it i.e.},
$F^{k_s\alpha_s} - \frac{1}{2} C_{i_p,\beta_p\;
j_q,\gamma_q}^{k_s,\alpha_s}\; A^{i_p,\beta_p} \wedge
A^{j_q,\gamma_q}$). Thus, the FDA's de Rham cohomology is trivial
and every closed form in the FDA may be written as $d$ of a form
constructed from its generators.

Let us now see how the above new gauge FDA's may be used to obtain
Chern-Simons (CS) gauge theories. Given a Lie algebra ${\mathcal
G}$, a Chern-Simons (CS) field theory is, generically, one for
which the Lagrangian form is the potential of a $2l$-form
$H=\langle F^{I_1},\dots, F^{I_l} \rangle$ constructed from the
curvature two-forms $F^I$ in the following way. Let $k_{I_1\dots
I_l}$ be a (graded)symmetric invariant tensor on ${\cal G}$, where
the index $I$ runs over the values of the ${\mathcal G}$ basis
index. Then, the CS Lagrangian  is a $(2l-1)$-form  $B$ such that
\begin{equation}
H=dB = \langle F^{I_1},\dots ,F^{I_l} \rangle= k_{I_1\dots I_l}
F^{I_1}\wedge\dots\wedge F^{I_l} \; ; \label{CSBa}
\end{equation}
as a result, the CS structure is intrinsically odd dimensional.
Since the {\it r.h.s.} is gauge invariant, the CS form $B$ is
gauge quasi-invariant {\it i.e.} its gauge transformation is given
by the differential of a $(2l-2)$-form\footnote{For Chern-Weil
invariants and for explicit expressions of CS forms and their
gauge transformation properties see {\it e.g.}, \cite{AI95}.}.
When $\cal G$ is a classical algebra (or one of its
$\mathbb{Z}_2$-graded counterparts), one may write $F=F^I T_I$,
$T_I$ being a matrix realization of the basis of $\cal G$. Then,
when different from zero, the (graded) symmetrized trace
sTr$(T_{I_1}\dots T_{I_l})$ gives a $G$-invariant symmetric
tensor, so that
\begin{equation}
     dB= \hbox{sTr} (F\wedge\dots\wedge F) \; .
     \label{CSBb}
\end{equation}
\noindent

    Consider the case of a CS theory for the algebra
${\mathcal G}(N_0,N_1,\ldots,N_n)$. Let $A^I$ and $F^I$ be
the gauge and curvature forms of the gauge FDA
associated with ${\mathcal G}=\oplus^n_0 V_p$, so that now
$I=i_1,\dots,i_n$ as in Th.~\ref{Th:newalg}. If $k_{I_1\dots I_l}$
is a (graded)symmetric invariant tensor of rank $l$ on $\cal G$,
the CS action associated with ${\mathcal G}$ is given by the
integral over a ($2l-1$)-dimensional manifold ${\cal M}^{2l-1}$ of
a potential form of $k_{I_1\dots I_l}F^{I_1}\wedge\dots\wedge
F^{I_l}$. Inserting in (\ref{CSBa}) the expansion of the gauge
forms ($\omega^{k_s,\alpha_s}\rightarrow A^{k_s,\alpha_s}$) one
finds\footnote{We shall ignore here the coupling constant and its
possible quantization (see \cite{DJT}). For the case {\it e.g.},
of odd-dimensional CS gravities, see \cite{Za95}; the quantization
may result from a mechanism similar to that associated with
Wess-Zumino-Witten-Novikov terms.}
\begin{equation}
\label{expCS}
         I[A,\lambda]=\int_{{\cal M}^{2l-1}}B(A,\lambda)=
           \int_{{\cal M}^{2l-1}}\sum^\infty_{\alpha=0} \lambda^N
B_N(A)=\sum^\infty_{\alpha=0}\lambda^N I_N [A]\ .
\end{equation}
For each order $N$, one obtains a CS action that is invariant
under the transformations (\ref{transf}), because it is the
integral of a form the differential of which is the coefficient of
$\lambda^N$ in the expansion of the invariant form
$H= \hbox{sTr} (F\wedge\dots\wedge F)$, and these coefficients are
separately invariant. Once an order $N$ is fixed, a gauge
FDA algebra is selected naturally: it is the one containing all
the gauge fields and curvatures that appear in $dB_N(A)$ in
agreement with Th.~\ref{Th:newalg} in Sec.~\ref{general}, since this
guarantees the consistency of (\ref{curvatures})-(\ref{Bianchi}).
In the case that we are considering, the fields $F^{k_s,\alpha_s}$
appear in the coefficient $dB_N(A)$ so that $N_p \le N$
for all $p$. The action $I_N[A]$ in (\ref{expCS}) could have
been obtained directly by using the corresponding symmetric
$G(N_0,\dots,N_n)$-invariant form.

Let us assume that $\cal G$ is simple. Then, with dimensionless
structure constants for $\mathcal{G}$, it is not possible to
assign consistently physical dimensions to its generators. The
fields $A^I$ are also dimensionless and the corresponding CS
integral cannot have dimensions of an action. One may, however,
rescale the fields $A^I$ as in a generalized W-W contraction (in
our scheme corresponding to $\mathcal{G}(0,1,\dots,n)$), $A^{i_p}
\rightarrow \lambda^p A^{i_p, p}$, and declare that $\lambda$ has
some definite physical dimensions. The resulting action is
constructed in terms of dimensionful fields, at the price of
introducing an explicit dependence of the structure constants on a
dimensionful parameter, which disappears by conveniently taking
the limit $\lambda\rightarrow 0$. This process gives a CS theory
on the contracted algebra when the gauge fields have suitable
physical dimensions. But the expansion method is more general, and
we may obtain true actions with the right physical dimensions, by
using (\ref{expCS}), due to the fact that  the action
$I_N [A]$ has dimensions $[\lambda]^{-N}$. Moreover, in
contrast with a contraction, the expansion method gives a CS
theory for a Lie (super)algebra that, in general, is of higher
dimension than that of ${\mathcal G}$. We now illustrate both
procedures, the contraction and the expansion one, using the case
of three-dimensional supergravity as an example.

\section{Application to Chern-Simons gauge theory of supergravity}
\label{sg}

It is well known that (super)gravities in three spacetime
dimensions are CS gauge theories \cite{PvN85,Paul86,Witten88} and
hence topological and exactly solvable \cite{Witten88}, which
allows for the construction of an exact quantum theory. For
instance, Poincar\'e supergravity in three spacetime dimensions is
a CS theory for the eight-dimensional $D$=3 superPoincar\'e Lie
algebra defined by the MC equations
\begin{eqnarray}
     d \sigma^{ab} &=&
-{\sigma^a}_c\wedge \sigma^{cb} \nonumber \\
d\pi^\alpha &=&
-\frac{1}{4}\sigma_{ab}{(\gamma^{ab})^\alpha}_\beta \wedge
\pi^\beta
\nonumber\\
d\Pi^a &=& -{\sigma^a}_b\wedge\Pi^b -\pi^\alpha
\gamma^a_{\alpha\beta} \wedge \pi^\beta \quad
(a,b=0,1,2\;,\;\alpha,\beta=1,2)\; , \label{spoin3}
\end{eqnarray}
where $\Pi^a$, $\pi^\alpha$ and  $\sigma_{ab}=-\sigma_{ba}$ are,
respectively, the translations, supertranslations and Lorentz MC
forms. The gauge FDA corresponding to (\ref{spoin3}), of gauge
one-forms $\omega_{ab}$, $\psi^\alpha$ and $e^a$ (corresponding,
respectively, to the MC forms $\sigma_{ab},\pi^\alpha,\Pi^a$) and
curvatures $R_{ab}$, ${\cal T}^\alpha$ and $T^a$, is
\begin{eqnarray}
        R_{ab} &=& d\omega_{ab}+
\omega_{ac}\wedge{\omega^c}_b\,:=\, D\omega_{ab} \quad, \nonumber\\
DR_{ab} &=& 0 \quad ; \nonumber\\
      {\cal T}^\alpha &=& d\psi^\alpha+\frac{1}{4} \omega_{ab}
{(\gamma^{ab})^\alpha}_\beta \wedge\psi^\beta := D\psi^\alpha
\quad,
\nonumber\\
      D{\cal T}^\alpha &=& \frac{1}{4} R_{ab}
{(\gamma^{ab})^\alpha}_\beta \wedge \psi^\beta \quad; \nonumber\\
          T^a &=& de^a + {\omega^a}_b\wedge e^b +\psi^\alpha
\gamma^a_{\alpha\beta} \wedge \psi^\beta := De^a +\psi^\alpha
\gamma^a_{\alpha\beta} \wedge \psi^\beta \quad; \nonumber  \\
     DT^a &=& {R^a}_b\wedge e^b+ 2{\cal T}^\alpha
\gamma^a_{\alpha\beta} \wedge \psi^\beta \quad . \label{Iso21g}
\end{eqnarray}
The action is then given by
\begin{equation}
        I= \int_{{\cal M}^3}\left(\epsilon^{abc}R_{ab}\wedge
e_c+ 4\psi_\alpha \wedge{\cal T}^\alpha \right),
      \label{poincareCS}
\end{equation}
where we take $\gamma^{abc} =\epsilon^{abc}$. The integrand is a
potential form of the closed gauge-invariant $4$-form
$\epsilon^{abc}R_{ab}\wedge T_c+4{\cal T}_\alpha \wedge {\cal
T}^\alpha$. Eq. (\ref{poincareCS}) is the action for Poincar\'e
supergravity in $2+1$ dimensions \cite{Gates} written using the
first order gauge formulation (see {\it e.g.}, \cite{Cd'AF91}),
the field equations of which are $T^a=0$ (equations for $\omega$;
metricity condition), $R^{ab}=0$ (equations for $e$; flat space
Einstein equations) and $\mathcal{T}^\alpha=D\psi^\alpha=0$
(equations for $\psi$ or three-dimensional counterpart of the
Rarita-Schwinger equations).

The superalgebra (\ref{spoin3}) can be viewed  as a contraction of
the Lie superalgebra $osp(1|2)\oplus osp(0|2) = osp(1|2)\oplus
sp(2)$, also known as the type (1,0) anti-de Sitter superalgebra
in $D=2+1$ \cite{Paul86}. Since the algebra is the direct sum of
two simple ones, no physical  dimensions can be assigned to the
forms. However one may rescale  some of them using a dimensionful
scale, which then appears explicitly in the structure constants,
and these rescaled generators have dimensions. From the
corresponding CS integral $I(\lambda)$ one may construct the
action $I(\lambda) / \lambda$, with dimensions of length (those of
an action in three dimensions in geometrized units), which is the
action for supergravity with a negative cosmological constant. The
limit $\lambda\rightarrow 0$ for $I(\lambda)/\lambda$ turns out to
be well defined, and gives (\ref{poincareCS}). Explicitly,
$osp(1|2)\oplus sp(2)$ may be given by the MC equations
\begin{eqnarray}
       d \sigma^{ab} &=&
-{\sigma^a}_c\wedge\sigma^{cb}-{\tilde \Pi}^a \wedge {\tilde
\Pi}^b- {\tilde \pi}^\alpha (\gamma^{ab})_{\alpha\beta} \wedge
{\tilde \pi}^\beta \nonumber
\\ d{\tilde \pi}^\alpha &=&
-\frac{1}{4}\sigma_{ab}{(\gamma^{ab})^\alpha}_\beta \wedge {\tilde
\pi}^\beta+\frac{1}{2} {\tilde \Pi}_a {(\gamma^a)^\alpha}_\beta
\wedge {\tilde \pi}^\beta
\nonumber\\
d{\tilde \Pi}^a &=& -{\sigma^a}_b\wedge{\tilde \Pi}^b -{\tilde
\pi}^\alpha \gamma^a_{\alpha\beta} \wedge {\tilde \pi}^\beta \quad
(a,b,c=0,1,2\;,\;\alpha,\beta=1,2) \quad, \label{adS3}
\end{eqnarray}
where, again, $\sigma^{ab}=-\sigma^{ba}$ and ${\tilde \Pi}^a$ are
bosonic and ${\tilde \pi}^\alpha$ is fermionic. Starting from this
algebra one may obtain the gauge FDA by using eq.
(\ref{FDAoriginal}) for the gauge forms ($\omega_{ab}$, ${\tilde
\psi}^\alpha$, ${\tilde e}^a$) and curvatures ($R_{ab}$,
$\tilde{\mathcal T}^{\alpha}$, ${\tilde T}^a$). These fields
cannot be assigned physical dimensions unless some of them are
rescaled. The obvious choice is to set ${\tilde \Pi}^a=\mu \Pi^a$,
${\tilde \pi}^\alpha=\mu^{1/2}\pi^\alpha$, and hence ${\tilde
e}^a=\mu e^a$, ${\tilde \psi}^\alpha=\mu^{1/2}\psi^\alpha$,
${\tilde T}^a=\mu T^a$, ${\tilde{\mathcal T}}^\alpha=
\mu^{1/2}\mathcal{T}^\alpha$  where the parameter $\mu$ has
dimensions $[\mu]=L^{-1}$ so that the algebra then reads
\begin{eqnarray}
       d \sigma^{ab} &=&
-{\sigma^a}_c\wedge\sigma^{cb}-\mu^2{\Pi}^a \wedge \Pi^b -
\mu\pi^\alpha (\gamma^{ab})_{\alpha\beta} \wedge \pi^\beta
\nonumber
\\ d\pi^\alpha &=&
-\frac{1}{4}\sigma_{ab}{(\gamma^{ab})^\alpha}_\beta \wedge
\pi^\beta+\frac{\mu}{2} \Pi_a {(\gamma^a)^\alpha}_\beta \wedge
\pi^\beta
\nonumber\\
d \Pi^a &=& -{\sigma^a}_b\wedge \Pi^b -\pi^\alpha
\gamma^a_{\alpha\beta} \wedge \pi^\beta \quad, \label{adS32}
\end{eqnarray}
and the contraction limit $\mu\rightarrow 0$ reproduces
(\ref{spoin3}). The associated gauge FDA
$(\sigma^{ab}\rightarrow\omega^{ab},\,\Pi_a\rightarrow e_a,\;
\pi^{\alpha} \rightarrow \psi^{\alpha})$ is
\begin{eqnarray}
        R_{ab} &=& d\omega_{ab}+
\omega_{ac}\wedge{\omega^c}_b +\mu^2 e_a \wedge e_b+
\mu\psi^\alpha
(\gamma_{ab})_{\alpha\beta} \wedge\psi^\beta \quad, \nonumber\\
DR_{ab} &=& \mu^2 T_a\wedge e_b-\mu^2 e_a\wedge T_b
+2\mu\mathcal{T}^\alpha (\gamma_{ab})_{\alpha\beta}
\wedge \psi^\beta \quad ;\nonumber\\
    {\cal T}^\alpha &=& d\psi^\alpha+\frac{1}{4} \omega_{ab}
{(\gamma^{ab})^\alpha}_\beta \wedge\psi^\beta-\frac{\mu}{2}
e_a{(\gamma^a)^\alpha}_\beta \wedge \psi^\beta := D\psi^\alpha
-\frac{\mu}{2} e_a{(\gamma^a)^\alpha}_\beta \wedge \psi^\beta
\quad ,\nonumber\\
      D{\cal T}^\alpha &=& \frac{1}{4} R_{ab}
{(\gamma^{ab})^\alpha}_\beta \wedge \psi^\beta-\frac{\mu}{2} T_a
{(\gamma^{a})^\alpha}_\beta \wedge \psi^\beta +
\frac{\mu}{2}e_a{(\gamma^a)^\alpha}_\beta \wedge \mathcal{T}^\beta
\quad ;\nonumber\\
          T^a &=& de^a + {\omega^a}_b\wedge e^b +\psi^\alpha
\gamma^a_{\alpha\beta} \wedge \psi^\beta := De^a +\psi^\alpha
\gamma^a_{\alpha\beta} \wedge \psi^\beta \quad ,\nonumber  \\
     DT^a &=& {R^a}_b\wedge e^b+ 2{\cal T}^\alpha
\gamma^a_{\alpha\beta} \wedge \psi^\beta\ .
                              \label{gaugeads3}
\end{eqnarray}
One may construct a three-dimensional CS theory starting from the
gauge invariant four-form $\epsilon^{abc}R_{ab} \wedge {\tilde
T}_c+4{\tilde {\cal T}}_\alpha\wedge{\tilde {\cal T}}^\alpha
=\mu\epsilon^{abc}R_{ab}\wedge T_c+4\mu{\cal T}_\alpha\wedge{\cal
T}^\alpha$. Since the gauge algebra is contractible, the CS
integral is then easily found to be
\begin{equation}
        I(\mu)= \mu\int_{{\cal
M}^3}\left(\epsilon^{abc}R_{ab}\wedge e_c+ 4\psi_\alpha\wedge
{\cal T}^\alpha-\frac{2}{3}\mu^2\epsilon^{abc}e_a\wedge e_b\wedge
e_c+2\mu\psi^\alpha(\gamma_a)_{\alpha\beta}\wedge \psi^\beta\wedge
e^a \right),
      \label{adsCS}
\end{equation}
which gives the $(1,0)$ $AdS$ supergravity lagrangian in differential
form, a supersymmetrization of $D=3$ gravity with negative cosmological
constant. Taking the $\mu\rightarrow 0$ limit in $I(\mu)/\mu$, the
CS supergravity action (\ref{poincareCS}) is recovered.

     It is worth clarifying at this stage the nature of the above
contraction. It is performed by writing $osp(1|2)\oplus sp(2)$ in
a (`pseudoextended') form that disguises its actual direct
(trivial) sum structure, which may be recovered by making the
change of basis
\begin{equation}
    \rho{'}^{\alpha\beta} = \frac{1}{4} \gamma_a^{\alpha\beta}
(\epsilon^{abc}\sigma_{bc}+2{\tilde \Pi}^a)\ ,\quad
    \rho^{\alpha\beta} = \frac{1}{4} \gamma_a^{\alpha\beta}
(\epsilon^{abc}\sigma_{bc}-2{\tilde \Pi}^a)\ , \quad \nu^\alpha =
\sqrt{2}{\tilde \pi}^{\alpha} \; , \label{changebasis}
\end{equation}
which exhibits the explicitly direct sum $osp(1|2)\oplus sp(2)$
form,
\begin{eqnarray}\label{explicitdir}
     d\rho^{\alpha\beta} &=&
-{\rho^\alpha}_\gamma\wedge \rho^{\gamma\beta}-\nu^\alpha\wedge
\nu^\beta \quad ,
\nonumber\\
     d\nu^\alpha &=& -  {\rho^\alpha}_\beta\wedge
\nu^\beta \quad ; \nonumber \\
     d\rho{'}^{\alpha\beta} &=&
-{\rho{'}^\alpha}_\gamma\wedge \rho{'}^{\gamma\beta} \quad
(\alpha,\beta=1,2)\; .
\end{eqnarray}
\noindent The contraction of (\ref{adS32}) does not respect the
above direct sum structure and, by not doing so, generates
non-trivial cohomology; this is why the $D=3$ superPoincar\'e
algebra may be obtained. This example is not unique. For instance,
some gauge formulations of $D$=1+1 gravity are based on a
four-dimensional central extension of the (1+1)-Poincar\'e algebra
with a `magnetic' modification of the momenta commutators. This
algebra is obtained by means of a so called `unconventional'
contraction \cite{CJ92,Riv94} of (the de Sitter or anti de Sitter
algebra) $so(2,1)$. This corresponds, actually, to making a
standard IW contraction of a trivial extension of $so(2,1)$ by a
one-dimensional algebra in such a way that the contracted algebra
becomes a non-trivial extension of the Poincar\'e one. Again, this
procedure corresponds to transforming \cite{Sal61,AA85} a
two-coboundary (that on $so(2,1)$, giving its trivial extension)
into a non trivial two-cocyle (on the ($1+1$)-Poincar\'e) by the
contraction limit (see {\it e.g.}, \cite{AI95} for the cohomology
that governs extension theory). Thus, these are all examples of
the first method mentioned in the Introduction, the contraction
one, which preserves the dimension of the algebra.

We now turn to the expansion method. Instead of using the
{\it eight}-dimensional $osp(1|2)\oplus sp(2)$ algebra to obtain
the $D$=3 superPoincar\'e by an IW contraction, we now take the {\it
five}-dimensional $osp(1|2)$ MC one as the starting point.
This superalgebra is also the de Sitter algebra in two dimensions.
Its MC eqs. are given by the first two equations in
(\ref{explicitdir}) (cf.~(\ref{ospmaurer})). We immediately see
that the $D=3$ superPoincar\'e algebra is $osp(1|2)(2,1)$, of
$\textrm{dimension}=2\textrm{dim}V_0+\textrm{dim}V_1=8$
(eq.~(\ref{eq:dimsuper})), making the identifications
$\rho^{\alpha\beta,0}=\frac{1}{4}(\gamma^{ab})^{\alpha\beta}\sigma_{ab}$,
$\nu^{\alpha,1}=\pi^{\alpha}$ and
$\rho^{\alpha\beta,2}=-\frac{1}{2} \Pi_a(\gamma^a)^{\alpha\beta}$
(the orders refer here to powers of $\lambda$, not $\mu$; note
also that in $D=3$ either $\gamma^a_{\alpha\beta}$ or their duals
$\gamma^{ab}_{\alpha\beta}$ provide a basis for the symmetric
tensors). The $osp(1|2)$ gauge FDA is generated by the gauge forms
$f^{\alpha\beta}$, $\xi^\alpha$ of curvatures
$\Omega^{\alpha\beta},\, \Psi^\alpha$, and is given by
\begin{eqnarray}
        \Omega^{\alpha\beta} &=& df^{\alpha\beta}+
{f^\alpha}_\gamma \wedge f^{\gamma\beta}+\xi^\alpha\wedge
\xi^\beta \nonumber \\
     d \Omega^{\alpha\beta} &=&
{\Omega^\alpha}_\gamma \wedge f^{\gamma\beta}- {f^\alpha}_\gamma
\wedge \Omega^{\gamma\beta}+\Psi^\alpha\wedge
\xi^\beta -\xi^\alpha\wedge \Psi^\beta \nonumber \\
    \Psi^\alpha &=& d\xi^\alpha +{f^\alpha}_\beta \wedge
\xi^\beta \nonumber\\
      d\Psi^\alpha &=& {\Omega^\alpha}_\beta \wedge
\xi^\beta -{f^\alpha}_\beta \wedge \Psi^\beta \quad
(\alpha,\beta=1,2) \quad , \label{osp21g}
\end{eqnarray}
where $f^{\alpha\beta}$, $\Omega^{\alpha\beta}$ are even and
symmetric in $\alpha,\beta$, and $\xi^\alpha$, $\Psi^\alpha$ are
fermionic. The indexes are raised and lowered by the $2\times 2$
antisymmetric matrix $\epsilon_{\alpha\beta}$,
$\xi_\alpha=\epsilon_{\alpha\beta}\xi^\beta$ and so on. The
corresponding gauge transformations, for parameters
$\Lambda^{\alpha\beta}=\Lambda^{\beta\alpha}$, $\varphi^\alpha$
corresponding to $f_{\alpha\beta}$ and $\xi^\alpha$ respectively
are
\begin{eqnarray}
         \delta f^{\alpha\beta} &=& d\Lambda^{\alpha\beta}
-{\Lambda^\alpha}_\gamma f^{\gamma\beta}+{f^\alpha}_\gamma
\Lambda^{\gamma\beta}-\xi^\alpha\varphi^\beta+
\varphi^\alpha\xi^\beta \nonumber     \\
\delta \xi^\alpha &=&
d\varphi^\alpha-{\Lambda^\alpha}_\beta\xi^\beta + {f^\alpha}_\beta
\varphi^\beta \quad. \label{transosp21}
\end{eqnarray}

    Let us now use the expansion method to obtain the CS
supergravity action in $D=2+1$ from a CS integral for $osp(1|2)$.
The CS integral based on (\ref{osp21g}) is constructed from the
gauge invariant closed four-form
\begin{equation}
\label{eq:csH} H={\Omega^\alpha}_\beta \wedge{\Omega^\beta}_\alpha
-2 \Psi_\alpha \wedge\Psi^\alpha \quad.
\end{equation}
We look for an integral form with dimensions of an action {\it
i.e.}, of length in geometrized units. If $\mu$ is the expansion
parameter, and $[\mu]=L^{-1}$, we need the order one in $\mu$ of
$H$. The present situation is that of Sec.~\ref{susy} with
$V^*_2=0$ and where $V_0^*$ and $V_1^*$ are generated by
$\rho^{\alpha\beta}$ and $\nu^\alpha$ respectively. Since $V_1$ is
a symmetric  coset, equation (\ref{eq:z2splitseries1}) for
$\lambda=\mu^{1/2}$ applies and  we have to consider the following
expansions
\begin{equation}
         f^{\alpha\beta}= \sum_{n=0}^\infty f^{\alpha\beta,n}
\mu^n \ ,\quad \xi^\alpha =\sum_{n=0}^\infty
\xi^{\alpha,n+1/2}\mu^{n+1/2} \quad , \label{exp3D}
\end{equation}
and similarly for the curvatures. We may obtain different FDA
gauge superalgebras by retaining different orders according to
Th.~\ref{Th:super} for $V_2=0$ (or to
(\ref{eq:Nzeroz2})-(\ref{eq:Nonez2})). On the other hand, the fact
that to construct the action we need the term proportional to
$\mu$ in the expansion of $H$ requires $n=1$, $n=0$ for the upper
limits of the sums in (\ref{exp3D}) (these correspond to the
$N_0=2,\,N_1=1$ that characterize $osp(1|2)(2,1)$), in agreement
with (\ref{eq:tress1}) and (\ref{eq:Nonez2}). So the relevant
algebra will correspond to the forms $f^{\alpha\beta,0}$,
$f^{\alpha\beta,1}$, $\xi^{\alpha,1/2}$,
\begin{equation}
    f^{\alpha\beta,0}=
\frac{1}{4}(\gamma^{ab})^{\alpha\beta}\omega_{ab}\ ,\quad
f^{\alpha\beta,1}= -\frac{1}{2}(\gamma^a)^{\alpha\beta}e_a\ ,\quad
\xi^{\alpha,1/2} = \psi^\alpha \quad . \label{osppoin}
\end{equation}
Then the resulting gauge FDA is precisely (\ref{Iso21g}), and the
term proportional to $\mu$ in the expansion of $H$ in
(\ref{eq:csH}) is the closed form
$\frac{1}{2}\gamma^{abc}R_{ab}\wedge T_c-2{\cal T}_\alpha\wedge
{\cal T}^\alpha$, the potential form of which leads to
(\ref{poincareCS}). This translates the fact that the $D=3$
superPoincar\'e algebra is $osp(1|2)(2,1)$.

The same procedure may be applied to obtain a CS theory based on
$osp(1|32)$ (see \cite{Ho99,Cham,Za00,Ba02,MoNi}), using either
the splitting of Th.~\ref{Th:super} or one with $V_2=0$, in which
case $V_0$ and $V_1$ are simply the bosonic and fermionic parts of
the superalgebra (as used at the end of Sec.~\ref{seis}). The
resulting algebras have a semidirect structure, where the Lorentz
($sp(32)$) algebra is the simple factor in the algebra resulting
from the first (second) splitting. Results on the corresponding
$D=11$ CS theory will be published separately.

\section{Conclusions and outlook}

In this paper we have described the expansion method, a procedure
of obtaining new (super)algebras ${\cal G}(N_0,\ldots,N_p)$ from a
given one ${\cal G}$ that we denote {\it expansions} of ${\cal G}$ 
(Th.\ref{Th:newalg}). It is based in the power expansion of the MC
equations that results from rescaling certain group variables.
These expansions are in principle infinite, but some truncations
are consistent and define the Maurer-Cartan equations of new
(super)algebras, the structure constants of which are obtained
from those of the original algebra ${\cal G}$. We have considered
the different possible  ${\cal G}(N_0,\ldots,N_p)$ algebras
subordinated to various splittings of ${\cal G}$ and discussed
their structure. We have seen that in some cases (when the
splitting of ${\cal G}$ satisfies the Weimar-Woods conditions) the
resulting algebras include the simple or generalized
\.In\"on\"u-Wigner contractions of ${\cal G}$, but that this is
not always the case. They may be, however, IW contractions of
certain higher-dimensional algebras related to the original one,
as we have seen in Sec.~\ref{sg}. In general, the new `expanded'
algebras have higher dimension than the original one.

Since ${\cal G}$ is the only ingredient of the expansion method,
it is clear that the extension procedure (which involves {\it two}
algebras) is richer when one is looking for new (super)algebras,
as discussed at the end of Sec.~\ref{seis}. As it is the case for
contractions, the expansion method is more constrained.
Nevertheless, we have used it to obtain the M-theory superalgebra,
including its Lorentz part, from $osp(1|32)$. After formally
extending the method to the case of gauge free differential
algebras, we have applied it to the case of CS supergravity in
$2+1$ dimensions where, using that $D$=3 superPoincar\'e is
$osp(1|2)(2,1)$, we have recovered the Chern-Simons supergravity
action from a CS form for $osp(1|2)$. The application of the
expansion method to the $D=11$ case will be presented elsewhere.

\medskip

{\bf Acknowledgments.} This work has been partially supported by
the Spanish Ministry of Science and Technology through grants
BFM2002-03681, BFM2002-02000 and EU FEDER funds, and by the Junta
de Castilla y Le\'on through grant VA085-02. Two of the  authors
also wish to thank the Spanish Ministry of Education and Culture
(M.P.) and the Generalitat Valenciana (O.V.) for their research
grants. A helpful discussion with I.~Bandos is also gratefully
acknowledged.

\appendix

\section{Appendix: expansion of $d\omega^{k_s,\alpha_s}$}
Inserting (\ref{eq:szero})-(\ref{eq:sn}) into (\ref{eq:MC}) where
now $p,q,s=0,1,\ldots,n$, and using
\begin{equation} \label{eq:sumatorio}
\left( \sum_{\alpha=p}^{\infty} \lambda^\alpha \omega^{i_{ p},
\alpha} \right) \wedge \left( \sum_{\alpha=q}^{\infty}
\lambda^\alpha \omega^{j_{ q}, \alpha} \right) =
       \sum_{\alpha=p+q}^{\infty}
\lambda^\alpha \sum_{\beta=p}^{\alpha-q} \omega^{i_{ p}, \beta}
\wedge \omega^{j_{ q}, \alpha-\beta} \; ,
\end{equation}
we obtain the expansion of the MC equations for ${\cal G}$,
\begin{equation} \label{eq:MCap}
\sum_{\alpha=s}^{\infty} \lambda^\alpha d\omega^{k_{ s}, \alpha}=
\sum_{\alpha=s}^{\infty} \lambda^\alpha \left[ -\frac{1}{2} c_{i_{
p}j_{ q}}^{k_{ s}} \sum_{\beta=0}^{\alpha} \omega^{i_{ p}, \beta}
\wedge \omega^{j_{ q}, \alpha-\beta} \right] \; ,
\end{equation}

\noindent
since the W-W conditions (\ref{eq:cont}) will give zero in the
r.h.s.~unless $\alpha=p+q \geq s$, in agreement with the l.h.s.
Eq.~(\ref{eq:MCap}) can be made explicit for $p,q,s=0,1,\ldots,n$
as follows: {\setlength\arraycolsep{2pt}
\begin{eqnarray}
\sum_{\alpha=s}^{\infty} \lambda^\alpha d \omega^{k_s , \alpha} &
= & -\frac{1}{2} \left[c_{i_0 j_0}^{k_s} \sum_{\alpha=0}^{\infty}
\lambda^\alpha \sum_{\beta=0}^\alpha \omega^{i_0 , \beta} \wedge
\omega^{j_0 ,\alpha-\beta} + 2 c_{i_0 j_1}^{k_s}
\sum_{\alpha=1}^{\infty} \lambda^{\alpha}
\sum_{\beta=0}^{\alpha-1} \omega^{i_0 , \beta} \wedge \omega^{j_1
,\alpha-\beta} + \ldots \right.
       \nonumber  \\
& & \left. +2 c_{i_0 j_n}^{k_s} \sum_{\alpha=n}^{\infty}
\lambda^{\alpha} \sum_{\beta=0}^{\alpha-n} \omega^{i_0 , \beta}
\wedge \omega^{j_n ,\alpha-\beta} +
       c_{i_1 j_1}^{k_s} \sum_{\alpha=2}^{\infty} \lambda^{\alpha}
\sum_{\beta=1}^{\alpha-1} \omega^{i_1 , \beta} \wedge \omega^{j_1
,\alpha-\beta} + \ldots \right.
\nonumber  \\
& & \left. +2 c_{i_1 j_n}^{k_s} \sum_{\alpha=1+n}^{\infty}
\lambda^{\alpha} \sum_{\beta=1}^{\alpha-n} \omega^{i_1 , \beta}
\wedge \omega^{j_n ,\alpha-\beta} + \ldots \right.
\nonumber  \\
& & \left. +c_{i_{n-1} j_{n-1}}^{k_s} \sum_{\alpha=2n-2}^{\infty}
\lambda^{\alpha} \sum_{\beta=n-1}^{\alpha-n+1} \omega^{i_{n-1} ,
\beta} \wedge \omega^{j_{n-1} ,\alpha-\beta} \right.
\nonumber  \\
& & \left. +2 c_{i_{n-1} j_{n}}^{k_s} \sum_{\alpha=2n-1}^{\infty}
\lambda^{\alpha} \sum_{\beta=n-1}^{\alpha-n} \omega^{i_{n-1} ,
\beta} \wedge \omega^{j_{n} ,\alpha-\beta} \right.
\nonumber  \\
& & \left. +c_{i_{n} j_{n}}^{k_s} \sum_{\alpha=2n}^{\infty}
\lambda^{\alpha} \sum_{\beta=n}^{\alpha-n} \omega^{i_{n} , \beta}
\wedge \omega^{j_{n} ,\alpha-\beta} \right] \quad .
\end{eqnarray}}

\noindent Rearranging powers we get {\setlength\arraycolsep{0pt}
\begin{eqnarray}
\sum_{\alpha=s}^{\infty} \lambda^\alpha d \omega^{k_s , \alpha}
&=& -\frac{1}{2} \left\{ c_{i_0 j_0}^{k_s} \omega^{i_0, 0} \wedge
\omega^{j_0 , 0} + \lambda \left[c_{i_0 j_0}^{k_s}
\sum_{\beta=0}^{1} \omega^{i_{0} , \beta} \wedge \omega^{j_{0}
,1-\beta} + 2 c_{i_0 j_1}^{k_s} \omega^{i_0, 0} \wedge
\omega^{j_1, 1} \right] \right.
\nonumber  \\
& &+ \lambda^2 \left[c_{i_0 j_0}^{k_s} \sum_{\beta=0}^{2}
\omega^{i_{0} , \beta} \wedge \omega^{j_{0} ,2-\beta} + 2 c_{i_0
j_1}^{k_s} \sum_{\beta=0}^{1} \omega^{i_0, \beta} \wedge
\omega^{j_1 , 2-\beta} \right.  \nonumber  \\
& &+ \left. \left. 2 c_{i_0 j_2}^{k_s} \omega^{i_{0} , 0} \wedge \omega^{j_{2} ,2}+
c_{i_1 j_1}^{k_s} \omega^{i_{1}, 1} \wedge \omega^{j_{1} ,1} 
\right] + \ldots \right\} \quad .
\label{eq:calc}
\end{eqnarray}

\noindent Eq.~(\ref{eq:calc}) now gives {\setlength\arraycolsep{0pt}
\begin{eqnarray}
\sum_{\alpha=s}^{\infty} \lambda^\alpha &d& \omega^{k_s , \alpha}
= -\frac{1}{2} c_{i_0 j_0}^{k_s} \omega^{i_0, 0} \wedge \omega^{j_0 ,
0}  \nonumber \\
-& & \sum_{\alpha=1}^{n-1} \lambda^\alpha \left[
       \frac{1}{2} \sum_{p=0}^{[\frac{\alpha}{2}]}
c_{i_p j_p}^{k_s} \sum_{\beta=p}^{\alpha-p} \omega^{i_p , \beta}
\wedge \omega^{j_p , \alpha-\beta} +
\sum_{p=0}^{[\frac{\alpha-1}{2}]} \sum_{q=p+1}^{\alpha-p} c_{i_p
j_q}^{k_s} \sum_{\beta=p}^{\alpha-q} \omega^{i_p , \beta} \wedge
\omega^{j_q , \alpha-\beta} \right]  \nonumber \\
-& & \sum_{\alpha=n}^{2n-1} \lambda^\alpha \left[
       \frac{1}{2} \sum_{p=0}^{[\frac{\alpha}{2}]}
c_{i_p j_p}^{k_s} \sum_{\beta=p}^{\alpha-p} \omega^{i_p , \beta}
\wedge \omega^{j_p , \alpha-\beta} +
\sum_{p=0}^{[\frac{\alpha-1}{2}]} \sum_{q=p+1}^{\textrm{min}
\left\{ \alpha-p \, , n \right\}} c_{i_p j_q}^{k_s}
\sum_{\beta=p}^{\alpha-q} \omega^{i_p , \beta} \wedge
\omega^{j_q , \alpha-\beta} \right]  \nonumber  \\
-& &\sum_{\alpha=2n}^{\infty} \lambda^\alpha \left[
       \frac{1}{2} \sum_{p=0}^{n}
c_{i_p j_p}^{k_s} \sum_{\beta=p}^{\alpha-p} \omega^{i_p , \beta}
\wedge \omega^{j_p , \alpha-\beta} + \sum_{p=0}^{n-1}
\sum_{q=p+1}^{n} c_{i_p j_q}^{k_s} \sum_{\beta=p}^{\alpha-q}
\omega^{i_p , \beta} \wedge \omega^{j_q , \alpha-\beta} \right] \, ,
\end{eqnarray}}

\noindent that is {\setlength\arraycolsep{0pt}
\begin{eqnarray}
\sum_{\alpha=s}^{\infty} \lambda^\alpha d \omega^{k_s , \alpha}
&=&
       -\frac{1}{2} c_{i_0 j_0}^{k_s} \omega^{i_0, 0} \wedge \omega^{j_0 ,
0} -
       \sum_{\alpha=1}^{\infty} \lambda^\alpha \left[ \frac{1}{2} 
\sum_{p=0}^
{\textrm{min} \left\{ [\frac{\alpha}{2}] \, , n \right\}} c_{i_p
j_p}^{k_s} \sum_{\beta=p}^{\alpha-p} \omega^{i_p , \beta} \wedge
\omega^{j_p , \alpha-\beta} + \right . \nonumber \\
& & + \left . \sum_{p=0}^{\textrm{min} \left\{
[\frac{\alpha-1}{2}] \, , n-1 \right\}} \sum_{q=p+1}^{\textrm{min}
\left\{ \alpha-p \, , n \right\}} c_{i_p j_q}^{k_s}
\sum_{\beta=p}^{\alpha-q}
       \omega^{i_p , \beta} \wedge \omega^{j_q , \alpha-\beta} \right] \; ,
\end{eqnarray}}

\noindent from which we obtain, upon explicit imposition of the
contraction condition (\ref{eq:cont}) on the structure constants
$c$'s:

\noindent $\alpha=s=0$:
\begin{equation}\label{facil}
d \omega^{k_0 , 0} = - \frac{1}{2} c_{i_0 j_0}^{k_0} \omega^{i_0 ,
0} \wedge \omega^{j_0 , 0} \; ;
\end{equation}
$\alpha =s \geq 1$, $s$ odd:
\begin{equation}
d \omega^{k_s , s} = -\sum_{p=0}^{\frac{s-1}{2}} c_{i_p
j_{s-p}}^{k_s} \omega^{i_p , p} \wedge \omega^{j_{s-p} , s-p} \; ;
\end{equation}
$\alpha=s \geq 1$, $s$ even:
\begin{equation}
d \omega^{k_s , s} = - \frac{1}{2} c_{i_{\frac{s}{2}}
j_{\frac{s}{2}}}^{k_s} \omega^{i_{\frac{s}{2}} , \frac{s}{2}}
\wedge \omega^{j_{\frac{s}{2}} , \frac{s}{2}}
-\sum_{p=0}^{\frac{s-2}{2}} c_{i_p j_{s-p}}^{k_s} \omega^{i_p , p}
\wedge \omega^{j_{s-p} , s-p} \; ;
\end{equation}
$\alpha > s \geq 0$:
\begin{eqnarray} \label{eq:MCexp2}
d \omega^{k_s , \alpha} = &-& \frac{1}{2} \sum_{p= \left[
\frac{s+1}{2} \right]}^{\textrm{min} \left\{ [\frac{\alpha}{2}] \,
, n \right\}} c_{i_p j_p}^{k_s} \sum_{\beta=p}^{\alpha-p}
\omega^{i_p , \beta} \wedge \omega^{j_p , \alpha-\beta} \nonumber \\
&-& \sum_{p=0}^{\textrm{min} \left\{ [\frac{\alpha-1}{2}] \, , n-1
\right\}} \sum_{q=\textrm{max} \left\{ s-p, p+1 \right\}}^
{\textrm{min} \left\{ \alpha-p \, , n \right\}} c_{i_p j_q}^{k_s}
\sum_{\beta=p}^{\alpha-q}
       \omega^{i_p , \beta} \wedge \omega^{j_q , \alpha-\beta} \; .
\end{eqnarray}

\end{document}